\documentclass[prd,aps,superscriptaddress,showpacs,amssymb,
amsmath,twocolumn,floatfix]{revtex4}
\usepackage{bm,graphicx}

\newcommand{\edinburgh}{School of Physics, University of
Edinburgh, King's Buildings, Edinburgh EH9 3JZ, U.K.}

\newcommand{\glasgow}{Department of Physics and Astronomy, 
University of Glasgow, Glasgow G12 8QQ, U.K.}

\newcommand{\cambridge}{Department of Applied Mathematics and
Theoretical Physics, University of Cambridge, Cambridge, U.K.}

\newcommand{\naive}{\textsc{One-link}}
\newcommand{\asqtad}{\textsc{Asqtad}}
\newcommand{\hyp}{\textsc{Hyp}}
\newcommand{\hisq}{\textsc{Fat7xAsq}}

\newcommand{\tr}{\mathbin{\text{Tr}}}
\newcommand{\la}{\left\langle}
\newcommand{\ra}{\right\rangle}

\newcommand{\msbar}{\overline{MS}}

\def\slashchar#1{\setbox0=\hbox{$#1$}           % set a box for #1 
   \dimen0=\wd0                                 % and get its size
   \setbox1=\hbox{/} \dimen1=\wd1               % get size of /
   \ifdim\dimen0>\dimen1                        % #1 is bigger
      \rlap{\hbox to \dimen0{\hfil/\hfil}}      % so center / in box
      #1                                        % and print #1
   \else                                        % / is bigger
      \rlap{\hbox to \dimen1{\hfil$#1$\hfil}}   % so center #1
      /                                         % and print /
   \fi}                                         %

\newcommand{\be}{\begin{equation}}
\newcommand{\ee}{\end{equation}}
\newcommand{\bea}{\begin{eqnarray}}
\newcommand{\eea}{\end{eqnarray}}
\newcommand{\D}{\slashchar{D}}

\begin{document}

\title{The Low-Lying Dirac Spectrum of Staggered Quarks}

\author {E. \surname{Follana}}
\affiliation{\glasgow}

\author{A. \surname{Hart}}
\affiliation{\edinburgh}

\author{\ ${}$C.T.H. \surname{Davies}}
\affiliation{\glasgow}

\author {Q. \surname{Mason}}
\affiliation{\cambridge}

\collaboration{${}$HPQCD and UKQCD collaborations}
\noaffiliation

\begin{abstract}
  
  We investigate and clarify the role of topology and the issues
  surrounding the $\varepsilon$-regime for staggered quarks. We study
  unimproved and improved staggered quark Dirac operators on quenched
  lattice QCD gluon backgrounds generated using a Symanzik-improved
  gluon action. For the improved Dirac operators we find a clear
  separation of the spectrum into would-be zero modes and others. The
  number of would-be zero modes depends on the topological charge as
  predicted by the continuum Index Theorem, and the expectation values
  of their chirality are large for the most improved actions
  ($\approx$ 0.7). The remaining modes have low chirality and show
  clear signs of clustering into quartets that become degenerate in
  the continuum limit. We demonstrate that the lattice spacing and
  volume dependence of the eigenvalues follow
  expectations. Furthermore, the non-zero modes follow the random
  matrix theory predictions for all topological charge sectors. The
  values of the chiral condensate extracted from fits to the
  theoretical distributions are consistent with each other, and with
  the results obtained from the total density of eigenvalues using the
  Banks-Casher relation. We conclude that staggered quarks respond
  correctly to QCD topology when both fermion and gauge actions are
  improved.

\end{abstract}

%\preprint{Edinburgh 2005-??}
%\preprint{GUTPA/05/??/??}

\pacs{11.15.Ha, % Lattice gauge theory
      12.38.Gc  % Lattice QCD
     }

\maketitle

\section{Introduction}
\label{sec_introduction}

The low energy regime of Quantum Chromodynamics (QCD) exhibits a rich
and interesting phenomenology, including the $U_A(1)$ axial anomaly,
chiral symmetry breaking and the topological properties of the theory.
A crucial step in elucidating these effects is understanding the
low-lying eigenvalue spectrum of the Dirac operator.  There are a
number of detailed predictions of the properties of these low-lying
modes, such as the existence of an Index Theorem, the Banks-Casher
relation, and the distribution of the first few eigenvalues in fixed
topological charge sectors.  Such effects are, however, inherently
non-perturbative and can only be studied fully using techniques like
lattice Monte Carlo simulation.

Universality decrees that any correct discretization of QCD must be
correct close enough to the continuum limit. If we want to make
phenomenological predictions, however, it is crucial that the QCD
lattice Dirac operator also has the correct low-lying spectrum at the
lattice spacings typically simulated.  By this we mean that the
deformations to the eigenvalue spectrum brought about by the
discretization must be small. In this case the relevant scale for
comparison is the light sea quark mass, $m_{u,d} = \mathcal{O}(5
\text{~MeV})$.

In this paper we show that these deformations are small enough for one
particular class of lattice fermions that are widely used in large
scale lattice simulations: improved staggered quarks.  Working at
similar lattice spacings in the quenched theory (i.e. pure Yang--Mills
gluodynamics) we show that the low-lying QCD Dirac eigenvalue spectrum
very closely reproduces the continuum features. In particular, we show
that improved staggered fermions do respond correctly to the gluonic
topological charge, and discuss why some confusion on this issue
exists in the literature.

To reach these conclusions we test a number of predictions,
quantitative and qualitative, for the low-lying modes: the
Atiyah--Singer Index Theorem, that relates the zero modes to the
topological charge of the gauge field; the Banks--Casher relation,
linking the chiral condensate to the density of eigenvalues at the
origin; the universality of the low-lying eigenvalue spectrum in the
$\varepsilon$-regime, and its connection to Random Matrix Theory. In
addition, we change both the lattice spacing and volume and show that
the variation of the spectrum matches theoretical expectations.

This paper expands and refines the work in
\cite{Follana:2004sz,Follana:2004wp}. 
We include much more extensive data, both on larger volumes and finer
lattices. Related work has been presented in \cite{Wong1,Wong2}.

The structure of the paper is as follows. In Sec.~\ref{sec_theory} we
describe our expectations for the Dirac eigenvalue spectrum, both in
the continuum and with lattice staggered fermions. We also review the
existing literature. We describe our methodology in
Sec.~\ref{sec_simulation}, and discuss the features of the spectrum in
Sec.~\ref{sec_spectrum}. In Sec.~\ref{sec_rmt} we compare our results
for the low-lying eigenvalues with the universal predictions from
random matrix theory.  We summarise and conclude our study in
Sec.~\ref{sec_summary}.

\section{Theoretical Background}
\label{sec_theory}

In this section we review the theoretical expectations for the
spectrum of the Dirac operator in the continuum and lattice staggered
cases.

\subsection{Continuum QCD}
\label{continuum}

A single, continuum fermion species is described by the massless,
gauge covariant Dirac operator, which is anti-Hermitian with a purely
imaginary eigenvalue spectrum:
\begin{equation}
\D f_s = i \lambda_s f_s \; , \qquad 
\lambda_s \in \mathbb{R} \;.
\end{equation}
Choosing orthonormalised eigenvectors, satisfying
\begin{equation}
f_s^\dagger~f_t = \delta_{st} \; , 
\end{equation}
the chirality of an eigenvector is defined by 
\begin{equation}
\chi_s \equiv f_s^\dagger \gamma_5 f_s \; .
\end{equation}
The Dirac operator also anticommutes with $\gamma_5$, 
\be
\{ \D,\gamma_5
\} = 0 \; ,
\ee
implying that the spectrum is symmetric about zero: 
\be
\mathop{\mathrm{sp}}(\D) = \{\pm i \lambda_s, \; \lambda_s \in \mathbb{R} \}
\; .
\label{eqn_evalue_pairs_continuum}
\ee If $\lambda_s \not = 0$, then $\gamma_5 f_s$ is also an
eigenvector with eigenvalue $-i \lambda_s$, and zero chirality,
$\chi_s = 0$. The zero modes, $\lambda_s = 0$, can be chosen with
definite chirality, $\chi_s = \pm 1$.
\subsubsection{Topological charge and the Index Theorem}
Smooth continuum gauge field configurations can be classified by their
winding number, which is an integer given by the expression:
\begin{equation}
Q = \frac{1}{32 \pi^2} \int d^4x \; \epsilon_{\mu \nu \sigma \tau}
\tr F_{\mu \nu}(x) F_{\sigma \tau}(x) \; .
\end{equation}
If we now consider the Dirac operator in the given gauge background,
and we denote by $n_{\pm}$ the number of zero modes with a given
chirality, their difference is related to the topological charge via
the Atiyah--Singer Index Theorem
\cite{Atiyah:1963,Atiyah:1968}:
\begin{equation}
Q = m \tr \frac{\gamma_5}{\D(A) + m} = \sum_s \chi_s = n_+ - n_- \; ,
\label{eqn_index}
\end{equation}
where $m$ is the quark mass (we keep the sign conventions of
Ref. \cite{Follana:2004sz}).

\subsubsection{The chiral condensate and the Banks--Casher relation}

The bare chiral condensate in a given gauge field background, $A$, is
given by the trace over the quark propagator:
\begin{equation}
\left\langle \overline{\psi} \psi(m) \right\rangle = - \frac{1}{V} \tr 
\frac{1}{\D(A) + m} \; .
\end{equation}
In the limit of $m \to 0$ it acts as an order parameter for the
spontaneous breaking of chiral symmetry. Expanding in eigenmodes of
$\D$ and using the $\pm i \lambda_s$ symmetry,
\begin{equation}
\left\langle \overline{\psi} \psi(m) \right\rangle = - \frac{1}{V} 
\sum_s \frac{1}{i\lambda_s + m} = - \frac{m}{V} 
\sum_s \frac{1}{\lambda_s^2 + m^2} \; .
\end{equation}
Averaging over a full ensemble of gauge fields, the sum over $s$ can
be replaced by an integral over eigenvalues with measure given by the
spectral density:
\begin{equation}
\rho(\lambda) \equiv \frac{1}{V} \sum_s \delta(\lambda - \lambda_s) \; .
\label{eqn_spectral_density}
\end{equation}
The integral has a smooth limit as $m \to 0$:
\begin{eqnarray}
\Sigma \equiv 
- \left\langle \overline{\psi} \psi (m=0) \right\rangle & = & 
\lim_{m \to 0} m \int_{-\infty}^\infty d \lambda \; 
\frac{\rho(\lambda)}{\lambda^2 + m^2} \nonumber \\
& = & \pi \rho(0) \; ,
\label{eqn_banks_casher}
\end{eqnarray}
which is the Banks--Casher relation. We expect the chiral condensate
$\Sigma$ to be non-zero when the chiral symmetry is spontaneously
broken. This requires that the large volume limit is taken before $m
\to 0$ to ensure that the discrete sum over eigenmodes can be replaced
by the smooth integral.

\subsubsection{The $\varepsilon$-regime and spectral universality}

The $\varepsilon$-regime of QCD occurs when the theory is
regularised in a finite volume such that the typical system size $L$
satisfies
\begin{eqnarray}
\mbox{$\varepsilon$-regime:} ~~~ &&
(\Lambda_{\text{QCD}})^{-1} \ll L \ll (M_\pi)^{-1} \; .
\label{eqn_rmt_hier}
\end{eqnarray}
In such systems, the volume is large enough that we may derive an
effective chiral field theory for pions by integrating out all other
hadronic states. The typical Compton wavelength of the particles so
removed is $(\Lambda_{\text{QCD}})^{-1}$. At the same time, the system
is very small compared to the Compton wavelength of the pions, so we
can further integrate out all their kinetic modes to leave us with a
theory describing just static pions.

It can then be rigorously shown that the remaining finite volume
partition function exactly corresponds to that of a random matrix
theory (RMT) with the same chiral symmetries as QCD
\cite{Damgaard:2001ep,Akemann:2003tv}.
This universality 
\cite{Leutwyler:1992yt,Shuryak:1993pi}
allows us to predict the distributions of low-lying non-zero
eigenvalues of the Dirac operator using RMT
\cite{Nishigaki:1998is,Damgaard:2000ah}.
The relevant universality class is determined by the chiral symmetries
of the finite volume partition function, and hence of QCD; for
fermions in the fundamental representation of the gauge group with
$N_c \ge 3$ the appropriate class is the chiral Unitary Ensemble (chUE)
with Dyson index $\beta=2$.

Rather than the absolute eigenvalue spacings, RMT predicts only the
\textit{relative} spacings of the eigenvalues in the form of
\textit{microscopic} (or \textit{unfolded}) spectral densities
$\rho_s(z)$. These detail the universal distribution of one or more
eigenvalues. There are separate predictions for each sector of
topological charge.  The absolute spectral densities include a
QCD-specific scale factor based on the chiral condensate:
\begin{equation}
\rho(\lambda) = \Sigma V \rho_s(\lambda \Sigma V) \; .
\label{eqn_unfolding}
\end{equation}
Comparison of the Dirac spectrum with RMT therefore provides a method
of determining the chiral symmetries of QCD, of probing the role of
topology, and of measuring the chiral condensate.

For the combined spectral density of all eigenvalues in a given
topological charge sector RMT predicts
\cite{Verbaarschot:1993pm}
\begin{equation}
\rho_{\text{all}}(z) = \frac{z}{2} \left[ J_{\nu}^2(z) - 
J_{\nu+1}(z)J_{\nu-1}(z) \right] \; ,
\label{eqn_rmt_all}
\end{equation}
where $J_{\nu}(z)$ are Bessel functions and $\nu = |Q|$. As these and
all other predictions depend only on $|Q|$, we improve the
statistical accuracy of the comparison with the Dirac spectrum by
combining the $\pm|Q|$ sub-ensembles.

There are also separate predictions for the $k^{\text{th}}$ eigenvalue
\cite{Forrester1993,Damgaard:2000ah}, e.g. $k=1$:
\begin{equation}
\rho_1(z) = \left\{
\begin{array}{l@{~~}l}
\frac{z}{2} e^{-z^2/4} \; , &
(\nu = 0) \; ,
\\ 
\frac{z}{2} I_2(z) e^{-z^2/4} \; , &
(\nu = 1) \; ,
\\ 
\frac{z}{2} \left( I_2^2(z) - I_3(z) I_1(z) \right) e^{-z^2/4} \; , &
(\nu = 2) \; .
\end{array}
\right.
\label{eqn_rmt_k}
\end{equation}
Expressions for higher $k$ are given in Ref.
\cite{Damgaard:2000ah}.
\subsection{Lattice staggered fermions}
\label{sec_stagg_discr}

The massless, gauge-invariant, \naive\ staggered Dirac operator on a
$d=4$ dimensional Euclidean lattice with spacing $a$ is
\begin{eqnarray}
\D (x,y) & = & \frac{1}{2au_0} \sum_{\mu=1}^d
\eta_\mu(x) \left[ U_\mu(x) \delta_{x+\hat{\mu},y} - H.c. \right] \;  \\
\eta_\nu(x) & = & (-1)^{\sum_{\mu < \nu} x_\mu }
\label{eqn_naive}
\end{eqnarray}
with $u_0$ an optional tadpole-improvement factor given, in our case,
by the fourth root of the mean plaquette. $\D$ is antihermitian and
obeys a remnant of the continuum $\gamma_5$ anticommutation relation:
\be
\left\{ \D,\epsilon \right\}  = 0 \;, \qquad 
\mathrm{with} \qquad
\epsilon(x)  = (-1) ^ {\sum_{\mu=1}^d x_\mu} \;.
\label{eqn_stag_symm}
\ee
As in the continuum case, its spectrum is therefore purely imaginary,
with eigenvalues occurring in complex conjugate pairs:
\be 
\mathop{\mathrm{sp}}(\D) =
\{\pm i \lambda_s, \lambda_s \in \mathbb{R} \} \; .
\label{eqn_evalue_pairs}
\ee
In fact, if $f_s$ is an eigenvector with eigenvalue $i \lambda_s$,
then $\epsilon \, f_s$ is also an eigenvector with eigenvalue $- i
\lambda_s$. 

The corresponding action describes $N_t = 2^{d/2}=4$ ``tastes'' of
fermions which interact via unphysical ``taste breaking'' interactions
that vanish in the continuum limit as $a^2$. In the limit $a \to 0$
there is an $SU(N_t) \otimes SU(N_t)$ chiral symmetry, and the
spectrum is therefore an exact $N_t$-fold copy of that described in
Sec.~\ref{continuum}.

At finite lattice spacing the chiral symmetry group is reduced to
$U(1) \otimes U(1)$ and we do not expect to see this picture, for
example there will not be an exact Index Theorem anymore. The relevant
operator for the Index Theorem must be a taste-singlet,
$\gamma_5^{ts}$, to couple to the vacuum correctly
\cite{Golterman:1984}. 
It is a point split operator in all $d$-dimensions, which can be made
gauge invariant by the insertion of the appropriate $U$ fields. It does
not anticommute with the staggered Dirac operator, $\{
\D,\gamma_5^{ts} \} \not = 0$. This also implies that the chirality of
the eigenmodes will not be exactly $\pm 1$ or $0$.

Exact zero eigenvalues only occur accidentally (and are completely
excluded at non-zero fermion mass), and all the eigenmodes will in
principle contribute in Eq.~(\ref{eqn_index})
\cite{Smit:1987fn}.
Due to the simple properties
\bea
[ \gamma_5^{ts}, \epsilon ] = 0 \; , \\
\epsilon^\dagger \epsilon = I \; ,
\eea
each pair of eigenvectors \{$f_s$, $\epsilon \, f_s$\} has the same
chirality $\chi_s$:
\be
\left\langle \epsilon f_s|\gamma_5^{ts} |\epsilon f_s
\right\rangle
 = \left\langle f_s| \epsilon^\dagger
\gamma_5^{ts} \epsilon |f_s \right\rangle = 
\left\langle f_s|\gamma_5^{ts}|f_s \right\rangle \; .
\ee
All of the preceding considerations depend only on simple symmetry
properties, and apply equally well to the improved staggered
operators we have used in our work.

Early studies of the \naive\ staggered operator and unimproved, Wilson
gauge action for coarse, thermalized lattices were done in
\cite{Smit:1987fn,Hands:1990wc,Venkataraman:1998yj,
  Hasenfratz:2003,Follana:2003}.
There was no clear separation in the low-lying modes, either in
eigenvalue or chirality.  The low-lying \naive\ eigenvalues also did
not follow the universal predictions for continuum QCD. In particular
they showed no dependence on the topological charge sector
\cite{Berbenni-Bitsch:1998tx,Damgaard:1998ie,
  Gockeler:1998jj,Damgaard:1999bq,Damgaard:2000qt}.  

In this study, however, we are interested in whether improvement of
the gauge action and Dirac operator allow the continuum--like spectrum
to be seen.

Certainly, some continuum features emerge using very smooth gauge
backgrounds (obtained by either repeated cooling or discretising
classical instanton solutions)
\cite{Smit:1987fn,Hands:1990wc}.
By contrast, here we are interested in thermalized, unsmoothed
configurations at lattice spacings (and gauge actions) typical of
lattice simulations, i.e. $a \approx 0.1$~fm.  We conjecture that
these failures of continuum predictions follow from the use of coarse
lattices and unimproved actions, resulting in large taste changing
interactions that ruin the continuum chiral symmetries.  We shall show
that the improved staggered operators introduced below allow all the
continuum features of the spectrum to emerge.

That improvement allows this to happen at such lattice spacings is not
entirely unexpected. QCD predicts that the topological susceptibility
is suppressed as we reduce the sea quark mass. This is seen in
simulations using improved gauge and staggered fermion actions
\cite{Hasenfratz:2001wd,Bernard:2002sa,Bernard:2003gq,Aubin:2004qz}.
With \naive\ staggered sea quarks no such variation is seen
\cite{Kogut:1990qd,Bitar:1991wr,Kuramashi:1993mv,Alles:2000cg,
  Hasenfratz:2001wd}.
Other evidence for the beneficial effect of improved staggered
operators was seen in the Schwinger model
\cite{Durr:2003xs,Durr:2004as,Durr:2004ab,Durr:2004rz,Durr:2004ta},
as well as in QCD
\cite{Follana:2004sz,Follana:2004wp,Durr:2004as,Durr:2004rz}.

\subsubsection{Improved staggered fermions}

\begin{figure}[b]
  \includegraphics*[width=2.9in]{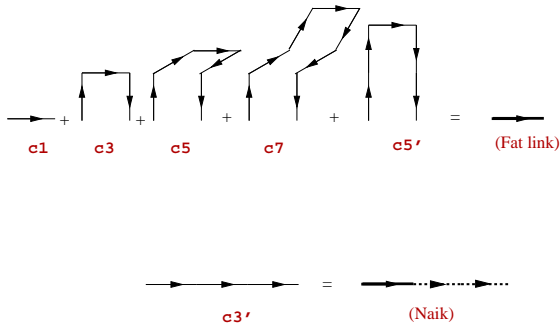}
  \caption{\label{fig_asqtad}Paths entering the improved staggered operators.}
\end{figure}

The continuum chiral symmetry is broken by the (unphysical)
taste-changing interactions. At leading order these involve exchange
of an ultraviolet gluon of momentum $q \approx \pi/a$.  Such
interactions are perturbative for typical values of the lattice
spacing. They can therefore be removed systematically using the method
of Symanzik.  This gives lattice operators with better taste symmetry
and smaller scaling violations. We do this by smearing the gauge field
to apply a form factor to the quark-gluon vertex and suppress the
coupling between quarks and high-momentum gluons
\cite{Lepage:1998vj,Bernard:1999xx,Orginos:1999cr}. 
This amounts to replacing the gauge link $U$ in Eq.~(\ref{eqn_naive})
by a ``fat'' link: a weighted sum of staples as shown graphically in
Fig.~\ref{fig_asqtad}. The choice of weight factors, $c_i$, determines
the degree of improvement. In this paper we compare three different,
improved operators known as \asqtad, \hisq\ and \hyp.

The first two are based on the \textsc{Fat7} operator which includes
staples of length up to $7a$. We can remove all $\mathcal{O}(a^2)$
discretization errors (at tree level) by adding two more terms to the
\textsc{Fat7} prescription, including a Naik three-link term
\cite{Naik:1986bn} and a five-link Lepage staple
\cite{Lepage:1998vj}. This operator is called \textsc{Asq}, whereas
\asqtad\ is the tadpole improved version which aims to also reduce the
radiative corrections to the scaling.

The improvement procedure may be iterated; for our most highly
improved operator the original links are first \textsc{Fat7} smeared
and projected back onto the $SU(3)$ gauge group. These fattened links
are then used to construct the \textsc{Asq} operator. We call this
operator \hisq. We stress that this is the same operator that was used
in
\cite{Follana:2004sz}.
Regrettably, there it was misnamed \textsc{Hisq}, a name
already used to describe a slightly different Dirac operator.

The final improved staggered Dirac operator is \hyp\
\cite{Knechtli:2000ku}.
Its construction is motivated by perfect action ideas, and involves
three levels of (restricted) APE smearing with projection onto $SU(3)$
at each level.  The restrictions are such that each fat link includes
contributions only from thin links belonging to hypercubes attached to
the original link.

The spectrum of all improved Dirac operators is of the form in
Eq.~(\ref{eqn_evalue_pairs}). Improved operators, however, show very
small taste-changing effects both in the hadronic splittings
\cite{Follana:2003, Follana:2004}
and, as we shall see, the eigenvalue spectrum.
\section{Details of the simulation}
\label{sec_simulation}

In this section we provide details of the gauge backgrounds used in
this study, and the methods used to determine the eigenvalues and the
topological charge. We also introduce notation for the eigenvalues to
be used in the following sections.

\subsection{The gauge action}

We use a quenched, $SU(3)$ gluonic action that is both tree level
Symanzik and tadpole improved
\cite{Curci,Curcierratum,Luscher,Luscher_erratum,Alford}:
\begin{equation}
S =
-\beta \sum_{{x \atop \mu<\nu}} \left( 
\frac{5}{3} P_{\mu \nu}(x) - \frac{1}{12} \frac{R_{\mu\mu\nu}(x)}{{u_0}^2}
-\frac{1}{12}\frac{R_{\mu\nu\nu}(x)}{{u_0}^2} \right) \; ,
\label{eqn_action}
\end{equation}
where $P$, $R$ are $1 \times 1$ and $2 \times 1$ Wilson loops
respectively. The tadpole improvement coefficient $u_0$ is defined as
the fourth root of the mean plaquette. 

We have generated five ensembles of around 1000 configurations, with
parameters for these ensembles as in Table~\ref{tab_ensembles} (taken
from Refs.~%
\cite{Bonnet:2001rc,Zhang:2001fk}).
The physical spatial size of the lattice is $aL$, and the volume $a^4
V \equiv a^4 L^3 T$. We have also used a few configurations from a
Wilson gauge ensemble with lattice spacing $a \approx 0.1$~fm and
volume $16^3 \times 32$.

Three ensembles have been tuned to study the effect of varying the
lattice spacing at fixed physical volume ($aL \approx 1.5$~fm).
Another three vary the volume at fixed $a=0.093$~fm.

The two coarser lattice spacings are representative of current, large
scale unquenched simulations using improved staggered quarks, with the
smallest $a$ being indicative of future calculations. Our gauge action
differs from the one used by the MILC Collaboration
\cite{Davies:2003ik}
only in small one loop radiative corrections which do not affect the
relevance of the results in this paper.

The main objective of this paper is to compare our results with the
expectations from the continuum limit and with the predictions from
RMT. For this purpose any systematic uncertainty in the determination
of the lattice spacings (for example due to finite-volume effects) is
not important. It would only become important for obtaining a precise
determination of some physical quantity. Therefore in general we
ignore this source of errors in our measurements, using only the
statistical errors based on a jack-knife analysis of the data.

\begin{table}[t]
\caption{Simulation parameters: the gauge coupling $\beta$, lattice 
spacing and volume and the spatial lattice extent.}
\label{tab_ensembles}

\begin{ruledtabular}
\begin{tabular}{cccc}
$\beta$ & $a$/fm & $V\equiv L^3 T$ & $aL$/fm \\
\hline
4.6 & 0.125 & $12^4$ & 1.50 \\
4.8 & 0.093 & $12^4$ & 1.12 \\
4.8 & 0.093 & $16^4$ & 1.49 \\
4.8 & 0.093 & $20^4$ & 1.86 \\
5.0 & 0.077 & $20^4$ & 1.54 \\
\end{tabular}
\end{ruledtabular}

\end{table}
\subsection{Measuring the topological charge}
\label{measuring_Q}

We measure the topological charge $Q$ using a number of gluonic
methods to look for consistency.  The topological charge of a gluonic
configuration is not uniquely defined at finite lattice spacing, and
different methods will sometimes disagree.  Over an ensemble, the
number of configurations that disagree on the value of $Q$ vanishes as
$a^2$ in the continuum limit.

We first cool the gauge fields and then apply a lattice topological
charge operator.  The cooling is (separately) done with two different
gauge actions (the Wilson and the 5Li
\cite{deForcrand:1995qq,deForcrand:1997sq}).
We measure the topological charge with the highly accurate ``5 Loop
improved'' (5Li) operator
\cite{deForcrand:1995qq,deForcrand:1997sq}. 
We then check that both charges are very close to integers and stable
under cooling. If the topological charges (rounded to the nearest
integer) defined by this procedure are the same for both actions, then
we assign the configuration to a definite topological charge sector.

Does this consistency criterion bias our fixed-$Q$ subensemble
averages? No; at $a=0.093$~fm ensembles fewer than $10\%$ of the
configurations have an ambiguous topological charge.  For the finest
ensemble, $a = 0.077$~fm, this number goes down to around $2\%$.

We have repeated parts of the analysis by ignoring the consistency
requirement and using only one of the cooling actions in the
definition of $Q$, assigning every configuration to a topological
charge sector.  The results are indistinguishable from those presented
here, and therefore this robustness criterion neither introduces a
systematic uncertainty in our results, nor changes the conclusions.

We stress that we only use cooling to determine the topological
charge. All the Dirac spectrum measurements are done on the original
thermalized (``hot'') configurations.

\subsection{Determining the eigenvalues}
\label{sec_num_methods}

To calculate the low-lying spectrum of the Dirac operators we use an
implementation of the Cullum-Willoughby Lanczos-based algorithm
\cite{Cullum_book}. Due to the structure of the staggered Dirac
operator the calculation of the spectrum can be done on an hermitian,
positive semi-definite operator, defined only on half of the lattice
(either the odd or the even sublattice). The algorithm can be extended
to the calculation of the corresponding eigenvectors, which are needed
to obtain the chirality.

On the ensembles used in this study we found no significant numerical
difficulties with this algorithm. In practice, the absence of exact
zero modes simplifies the calculation from a numerical point of view,
as the linear operator we are dealing with is then positive definite.

Due to the exact $\pm$ symmetry, we only need to consider half of the
spectrum, say $\{i \lambda_s, \, \lambda_s > 0\}$ (which we call the
positive sector). The other half is a mirror image which we call the
negative sector. From now on we dispense with the imaginary unit $i$,
and discuss and show on the plots the imaginary part of the positive
half spectrum $\{\lambda_s, \, \lambda_s > 0\}$.

For the analysis we divide the (positive) eigenmodes on each
configuration into two sets based on the gluonic topological charge.
Inspired by our continuum limit expectations, we expect the $4 |Q|$
near-zero modes to divide into $2|Q|$ modes each side of zero. We call
the arithmetic mean of the positive near-zero modes $\Lambda_0$.

The remaining positive modes are divided into quartets. We call the
means of successive sets $\Lambda_{1,2,\dots}$. We define the
intra-quartet splitting of the $s^\text{th}$ set, $\delta \Lambda_s$,
as the difference between the largest and smallest eigenvalues in the
quartet. We also define the inter-quartet gap between the
$s^\text{th}$ and $(s+1)^\text{th}$ quartets, $\Delta \Lambda_s$, as
the difference between the smallest eigenvalue of the higher quartet
and the largest eigenvalue of the lower quartet.

We use the notation $\langle~\cdot~\rangle_Q$ to denote the
expectation value over sub-ensembles with fixed gluonic topological
charge $\pm Q$.

Unless otherwise specified, the eigenvalues are expressed in physical
units of MeV, with the scale set using the quoted lattice spacing.

\section{Spectrum and Index Theorem}
\label{sec_spectrum}

In this section we study the Dirac spectra, and show how the continuum
results emerge for the improved operators.

\subsection{The Index Theorem}

\begin{figure*}[t]
\includegraphics[width=7in]{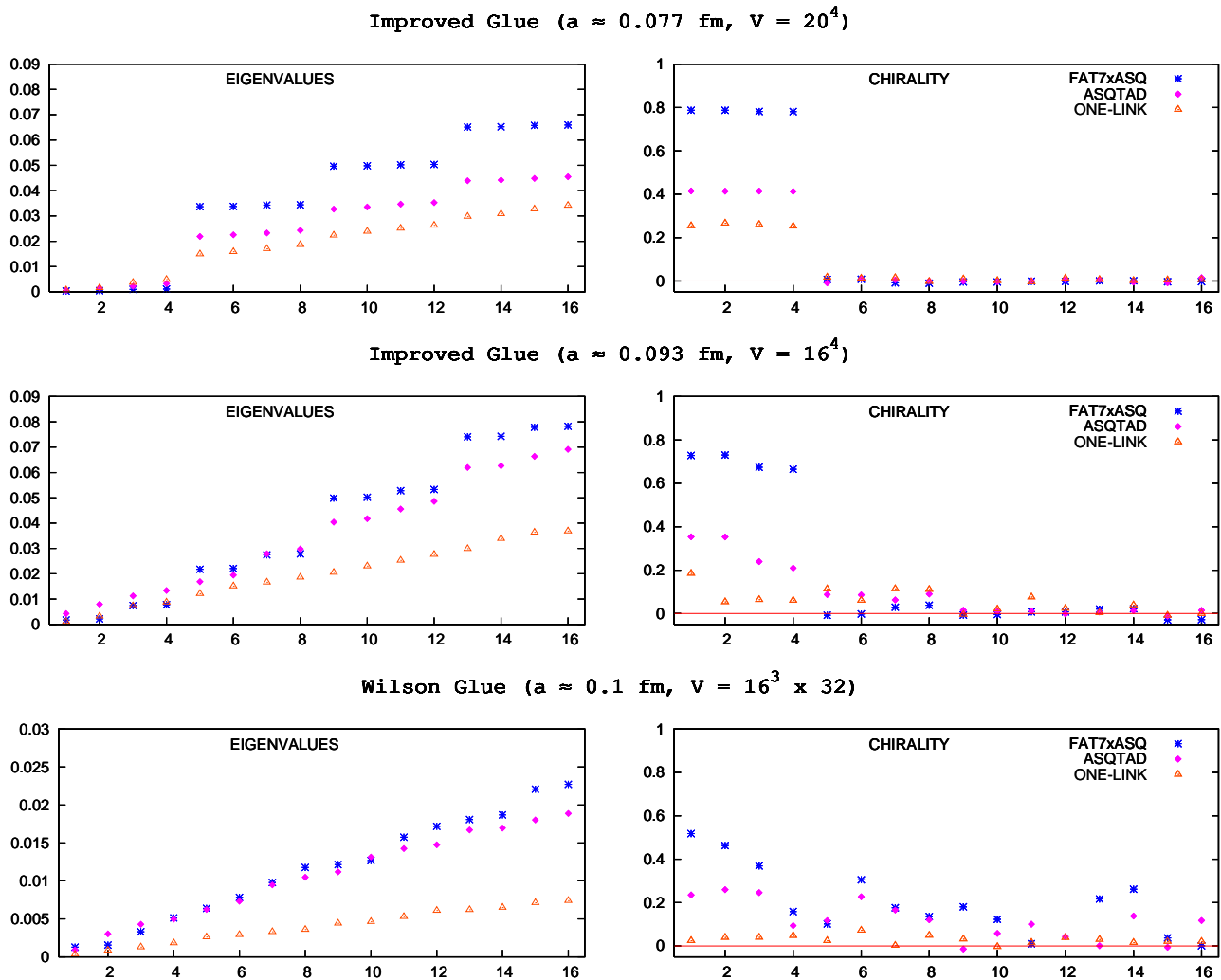}

\caption{\label{fig_spectrum} The positive half of several typical
  low-lying eigenmode spectra for configurations of $Q = 2$, in
  lattice units. The $x$-axis is eigenvalue number. The \hyp\ operator
  gives results very similar to \hisq, and they are not shown for
  clarity.}
\end{figure*}

We begin our analysis by qualitatively comparing the low-lying modes
of various staggered quark operators on typical gauge backgrounds
selected from those with $Q~=~2$.  Working at approximately fixed
lattice volume ($aL \approx 1.50$~fm), we compare three different
gauge field actions: unimproved Wilson at $a=0.1$~fm; Symanzik
improved, Eq.~(\ref{eqn_action}), at a comparable $a=0.093$~fm, and at
the finer $a=0.077$~fm. In Fig.~\ref{fig_spectrum} we show both the
value and the chirality of the first sixteen positive eigenvalues.

Near the continuum limit we expect to see first $2|Q|=4$ near-zero
modes with their chirality renormalised slightly away from unity. The
remaining modes should have chirality near zero and divide into almost
degenerate quartets with intra-quartet splitting much
smaller than the inter-quartet gap.

For the improved gauge action on the finer lattice the spectrum looks
quite continuum-like for all Dirac operators and we see a clear Index
Theorem. For the improved operators we also see a very clear quadruple
degeneracy in the non-zero modes. The renormalisation of the chirality
away from 1 is small for the improved Dirac operators (around
$Z=1.2$), becoming as large as $Z \approx 4$ for the \naive\
fermions. This result is fairly typical of operator renormalisation
for staggered quarks, with the improved actions showing significantly
better renormalisation factors \cite{Hein,Trottier:2002,Lee:2002}. We
will discuss later the renormalisation of the chiral condensate and
see the same effect.

On the coarser improved gauge background, we still have a good
approximation to the Index Theorem using the \hisq\ and \hyp\
operators: there is a separation between near-zero modes of high
chirality and the rest of the spectrum; the number of near-zero modes
is $2|Q|$; and all of them have very similar chiralities. The rest of
the spectrum have almost zero chirality and cluster into quartets.
These properties can be seen to a lesser extent for the \asqtad\
operator, but are absent in the \naive\ case.

Nothing clear can be seen in the unimproved, Wilson gauge case.
Improving the Dirac operator makes a difference, increasing the
chirality of the low modes. Nonetheless the chiralities do not show a
clear Index Theorem, and the eigenvalues do not gather into quartets.

\begin{figure}[t]
\includegraphics[width=3.5in,clip]{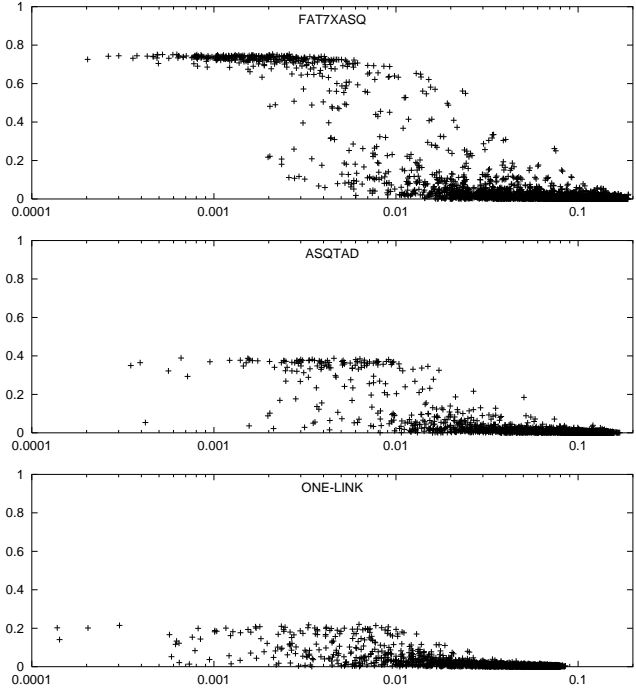}

\caption{\label{fig_scatter1} A scatter plot for different staggered
quark formulations, with absolute value of the chirality on the y axis
and eigenvalue $\lambda_s$ on the x axis, in lattice units. The lowest
50 eigenvalues for 147 configurations from our $16^4$, $a$=0.093~fm
ensemble are plotted. }
\end{figure}

\begin{figure}[t]
\includegraphics[width=3.5in, height=1.5in, clip]{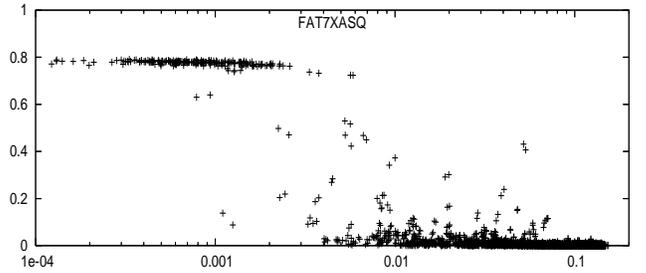}

\caption{\label{fig_scatter2} A scatter plot for \hisq, with absolute
value of the chirality on the y axis and eigenvalue $\lambda_s$ on the
x axis, in lattice units. The lowest 50 eigenvalues for around 100
configurations from our $20^4$, $a=0.077$~fm ensemble are plotted. }
\end{figure}

To give an idea of the behaviour over a large ensemble,
Fig.~\ref{fig_scatter1} shows a scatter plot of the absolute value of
the chirality versus the (positive) eigenvalues. Results for different
operators are plotted, using the same $a =0.093$~fm ensemble of gauge
fields. We can see the formation of a gap between modes of small and
large chirality as we improve the staggered operators, as well as the
overall increase in chirality of the near-zero modes. As seen in
Fig.~\ref{fig_scatter2} for the most improved operator on the finest
ensemble, the separation gets sharper as we approach the continuum
limit.

The separation is not strict even for the most improved action, with a
few configurations showing intermediate values of the chirality. At
least some of these cases are associated with configurations where the
$\mathcal{O}(a^2)$ ambiguity in the topological charge is large and/or
where the total number of near-zero modes is greater than the minimum
prescribed by the Index Theorem, i.e. $n_+ + n_- > N_t |Q|$.


As the near-zero modes for the more improved actions clearly separate
and have well defined chirality, we may define a fermionic index
$\overline{Q} = {(n_+ - n_-)}/{N_t}$ on each
configuration. $\overline{Q}$ is then strongly correlated with $Q$,
and we have a very good approximation to the Index Theorem. We expect
the number of configurations where $Q \not= \overline{Q}$ to vary as
$\mathcal{O}(a^2)$. Although we don't have enough statistics to test
this scaling quantitatively, we can get some estimates from our
results.  For our $a = 0.093$~fm ensemble, the disagreement between
$Q$ and $\overline{Q}$ is around $10\%$, and it goes down to about
$2\%$ on the finest $a = 0.077$~fm ensemble. This ambiguity is
therefore of the same order as the one intrinsic in the gluonic
definition, as discussed in Sec.~\ref{measuring_Q}. We stress
here that $\overline{Q}$ played no part in the assignment of
configurations into sectors of fixed topological charge.

We may worry that when $Q \not= \overline{Q}$ the division of the
eigenvalues into near-zero and non-zero modes, and the division of the
latter into quartets, will be skewed. For instance, on such
configurations the intra-quartet splitting will be as large as the
inter-quartet gap. The examination of the histograms of $\delta
\Lambda_s$ and $\Delta \Lambda_s$ show that in all cases any slight
bias on the mean is swamped by the statistical spread.

\subsection{The Banks--Casher relation}
\label{sec_Banks-Casher}

We begin our quantitative analysis with the Banks--Casher relation
(\ref{eqn_banks_casher}), which links the density of non-zero
eigenvalues at the origin with the chiral condensate, and thus
provides a model independent determination of~$\Sigma$. For $N_t$
degenerate flavours, the Banks-Casher relation would be modified to
\begin{equation}
\Sigma = \pi \rho(0) / N_t \; .
\label{eqn_banks_casher_Nt}
\end{equation}
This is the relevant form for staggered fermions, with $N_t = 4$. Note
that the appropriate $\overline{\psi}\psi$ operator for the chiral
condensate is the taste-singlet, but this is the local operator and so
there are no other modifications to the Banks-Casher relation for
staggered quarks.

\begin{figure}[t]
\includegraphics[width=3.3in, clip]{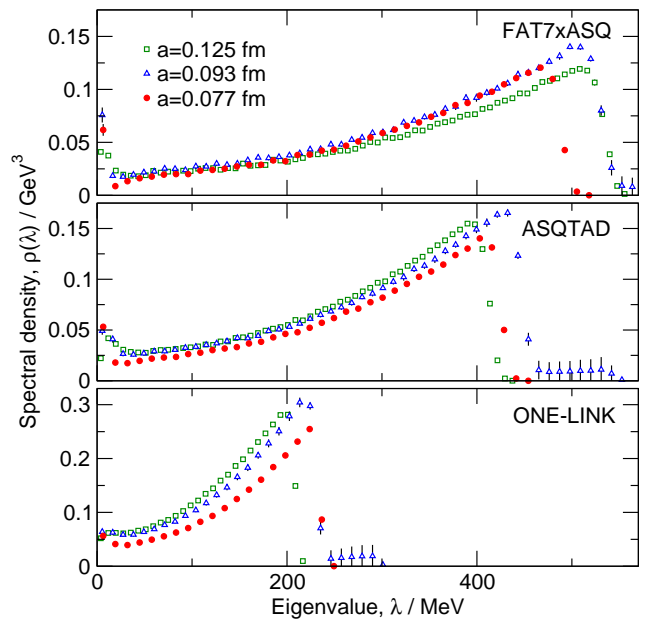}

\caption{\label{fig_spec_dens} The spectral density for different
  Dirac operators at fixed lattice volume $aL=1.5$~fm. Only the
  contribution of the first 100 eigenvalues have been calculated.}
\end{figure}

In Fig.~\ref{fig_spec_dens} we plot a histogram of the spectral
density averaged over all configurations in the ensemble. The errors
come from a jack-knife analysis of the data.

The spectrum divides clearly into two separate sections: a sharply
peaked spike near $\lambda=0$, coming from the near-zero modes, and a
broader distribution at larger eigenvalue from the remaining
modes. This separation is much sharper for the improved actions. The
spectral densities corresponding to the improved actions also show an
excellent scaling behaviour with the lattice spacing. 

For larger $\lambda$, a $\lambda^3$ behaviour is expected to set
in. Having only calculated the first 100 eigenvalues, however, we
instead find the spectral density cut off at larger $\lambda$. It
would be necessary to include many more modes to see this behaviour.

%Over the majority of the range the density of non-zero modes varies
%linearly with $\lambda$.  At small $\lambda$, the spectral density is
%suppressed. As the volume increases, the gap between the breakdown of
%the linear region and the zero mode peak becomes smaller.

%The gap is expected to vary as $V^{-1/2}$ 
%
%\cite{Hands:1990wc}.
%
%It is difficult to be quantify this from our data, but we do see broad
%agreement with this across the volumes available at $\beta=4.8$. For
%instance, with the \hisq\ operator the gap appears below 125, 100 and
%55~MeV respectively for $L=12,~16,~20$.
%
%
%
%
%
\begin{table}[t]

\newcommand{\noentry}{\multicolumn{1}{c}{---}}

\caption{\label{tab_rmt_chir_cond} Estimates of the cube root of the
bare chiral condensate, $\Sigma^{1/3}$, (in units of MeV) from: the
Banks-Casher relation (upper section) and comparisons of eigenvalue
distributions with random matrix theory (lower section).  }

\begin{ruledtabular}
\begin{tabular}{ccllll}
& &\multicolumn{4}{c}{Dirac operator} \\
\cline{3-6}
$a$/fm & $aL$/fm & \naive\ & \asqtad\ & \hyp\ & \hisq\ \\
\hline
0.125 & 1.50 & 348~(10) & 274~(10) & \noentry & 256~(10)  \\
0.093 & 1.12 & 330~(30) & \noentry & \noentry & 260~(10) \\
0.093 & 1.49 & 349~(10) & 275~(10) & 279~(10) & 260~(10) \\
0.093 & 1.86 & 360~(10) & \noentry & \noentry & 255~(10) \\
0.077 & 1.54 & 322~(10) & 253~(10) & \noentry & 238~(10) \\
\hline \hline
0.125 & 1.50 & 356~(8)  & 283~(6)  & \noentry & 260~(6)  \\
0.093 & 1.12 & 383~(43) & \noentry & \noentry & 289~(33) \\
0.093 & 1.49 & 361~(12) & 293~(5)  & 279~(4) & 281~(2) \\
0.093 & 1.86 & 358~(6)  & \noentry & \noentry & 286~(4) \\
0.077 & 1.54 & 321~(10) & 266~(10) & \noentry & 256~(5) \\
\end{tabular}
\end{ruledtabular}
\end{table}

Using Eq.~(\ref{eqn_banks_casher_Nt}), we can extract the single taste
chiral condensate by extrapolating the density of non-zero modes to
zero, $\rho(\lambda) \rightarrow \rho(0)$.  The chiral condensate is
the order parameter for chiral symmetry breaking, and we see clear
evidence from Fig.~\ref{fig_spec_dens} that all our lattices are large
enough that $\Sigma \not= 0$.

To estimate $\rho(0)$ we carry out uncorrelated fits to
$\rho(\lambda)$, excluding the near-zero mode peak from the fit
range. We use a polynomial fit function
\begin{equation}
\rho(\lambda) = \rho_0 + \lambda \rho_1 + \lambda^2 \rho_2 + \lambda^3 \rho_3
\; .
\end{equation}
We carry out a number of fits to each data set, constraining different
combinations of $\rho_i$ to zero and varying the fit range. The
intercept $\rho_0$ is then used to obtain the chiral condensate values
given in the upper half of Table~\ref{tab_rmt_chir_cond}. The quoted
errors cover the fit function and range uncertainties, which dominate
the statistical and fitting errors.

Note that the \naive\ density at the origin $\rho(0)$ is much larger
than the one for the improved case. This is clearly noticeable when
inverting the operator, with the improved actions needing fewer
iterations than the \naive\ action (also seen in
\cite{Orginos:1999cr}).

The larger $\rho(0)$ gives a much higher value for $\Sigma$ for the
\naive\ action, but it is important to note that this is the bare
lattice chiral condensate and does not have to agree between actions
in the same way that the bare lattice quark mass does not. What should
agree is the chiral condensate defined in some physical way, for
example in the $\overline{MS}$ scheme at a fixed scale. The
appropriate renormalisation factor for $\overline{\psi}\psi$ is the
inverse of that for the quark mass and this has been calculated to
$\cal{O}$$(\alpha_s)$ for the staggered quark actions used here and
the improved gluon action (at $\cal{O}$$(\alpha_s)$ there is no effect
from one-loop improvements to the gluon action or the presence of sea
quarks). The quark mass renormalisation was quoted in
\cite{Aubin:2004mas} as:
\begin{eqnarray}
  m^{\msbar}(\mu)\! & = &   \nonumber \\ 
 \!\frac{am_0}{a} && \hspace{-6mm} 
\left(1\!+\!\alpha_V(q^*)\,Z_m^{(2)}\!\left(a\mu,(am)_0\right)
+ \mathcal{O}(\alpha^2)\right),   \nonumber \\ 
Z_m^{(2)}(a\mu,am_0) & = & 
\left(b(am_0)-\frac4{3\pi}-\frac2\pi\ln(a\mu)\right).
\end{eqnarray}
where $b(am_0)$ is the finite piece of the lattice mass
renormalisation and has small $am_0$ dependence. The $\frac{4}{3\pi}$
comes from the continuum mass renormalisation. With
tadpole-improvement $b(0)$ takes the value 0.543 for the \asqtad\
quark action and 3.6 for the \naive\ action (denoted naive staggered
quarks in \cite{Aubin:2004mas}). For \hisq\ $b(0)$ is 0.375. For the
chiral condensate renormalisation, $Z_S$, we need to take the inverse
of the renormalisation above. We take $\mu$ to be 2 GeV, as a standard
reference point, and take the scale of $\alpha_s$ to be 2/a, as
determined for the \asqtad\ action at comparable values of $a$ in
\cite{Aubin:2004mas}. The values of $\alpha_s$ in the $V$ scheme in
the quenched approximation are taken from \cite{Mason}. It is then
clear that $Z_S$ for the \asqtad\ and \hisq\ actions is very close to
1 for all lattice spacing values, but significantly smaller than 1 for
the \naive\ action.  The renormalisation brings the value of the
physical chiral condensate for the \naive\ action from the
Banks-Casher relation down to 290 MeV, much closer to that from the
improved actions. The renormalisation constant is so far from 1 in
this case, however, that significant higher order corrections are to
be expected.  For the improved actions, however, the renormalisation
is very small indeed and numbers change by less than 2\% from the bare
results of table \ref{tab_rmt_chir_cond}. The estimated uncertainty
from higher order perturbative corrections is 4\% i.e. $1\times
\alpha_s^2$. From table \ref{tab_rmt_chir_cond} we also have a
statistical error of 4\%. In addition there are statistical and
systematic errors from the determination of the lattice spacing that
are relevant here. All of these errors are small compared to the
overall 10-20\% error from the inability to determine the lattice
spacing uniquely in the quenched approximation. This means that any
physical value for the chiral condensate that we quote will have a
significant error from this source alone. Our results for the \hisq\
action give a result for $\left(- \langle
\overline{\psi}\psi\rangle^{1/3}\right)^{\overline{MS}}(2 \rm{GeV})$
of 260(10)(10)(30) MeV, where the first error is statistical, the
second perturbative and the third a quenching error. This is in
agreement with other results in the literature in the quenched
approximation (for example, \cite{Giusti:1998}). An improvement on
this value requires unquenched configurations, but our result shows
that an accurate calculation is certainly possible with improved
staggered quarks.

\subsection{Scaling of the near-zero modes}

The near-zero eigenvalues for a given Dirac operator vary both with
the lattice spacing and volume.

\begin{figure}[t]
\includegraphics[width=3in, clip]{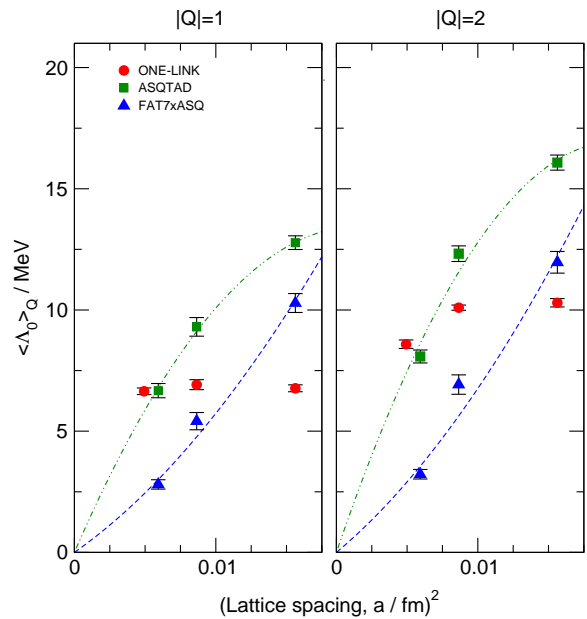}

\caption{\label{fig_zero_scal} The scaling of the near-zero modes with
  lattice spacing at fixed lattice volume for different topological
  charge sectors. The near-zero modes are averaged. Also shown are
  linear plus quadratic fits through the origin for \asqtad\ and
  \hisq\ .}
\end{figure}

In Fig.~\ref{fig_zero_scal} we show the variation of the mean
near-zero modes $\la \Lambda_0 \ra$ with lattice spacing at fixed
physical volume ($aL \approx 1.5$~fm). The panels show the results for
sectors of topological charge $|Q|=1$ and $2$.
For a sufficiently small lattice spacing, we expect the eigenvalues to
go to zero as $(a^2)$ plus $\mathcal{O}(a^4)$ corrections. The
improved staggered results show this trend across the full range of
lattice spacings studied, with small $\mathcal{O}(a^4)$ corrections,
slightly positive for \hisq\ but negative for \asqtad. By contrast,
the \naive\ near-zero modes show very strong deviations from the
leading order scaling behaviour.

In almost all quantities that we measure the \hyp\ operator gives
results that agree with \hisq\ within one standard error. Unless there
is a significant difference we do not discuss the \hyp\ data and for
clarity they are not shown on the figures, but we have checked that
all statements regarding \hisq\ apply equally to \hyp.

The \hisq\ near-zero modes are numerically smaller than those of the
\naive\ action for $a \lesssim 0.13$~fm. The \asqtad\ near-zero modes
are smaller than the \naive\ ones for $a \lesssim 0.08$~fm. Even when
the modes are similarly sized, however, we have seen earlier that
there is a qualitative difference in the chirality as we improve the
action. This is a clear signal that the operator improvement is
working.
\begin{figure}[t]
\includegraphics[width=3in, clip]{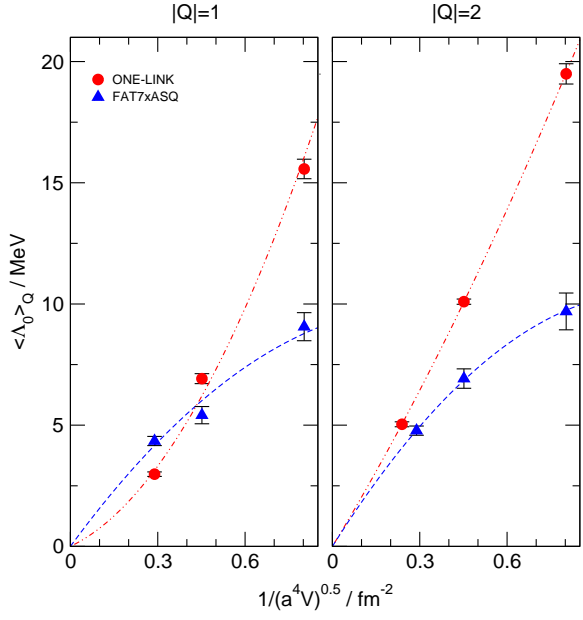}

\caption{\label{fig_zero_vol} The averaged near-zero modes as a
  function of lattice volume at $a=0.093$~fm for different topological
  charge sectors. Also shown are linear plus quadratic fits through
  the origin. }
\end{figure}

The variation of the near-zero eigenvalues with volume at fixed
lattice spacing is shown in Fig.~\ref{fig_zero_vol}. $\la \Lambda_{0}
\ra$ is expected to vary as the inverse square root of the volume.
This can be justified in a semiclassical picture, where the
topological charge density is dominated by a dilute gas of instantons.
The finite volume corrections to the zero modes come from the lack of
large instantons, excluded from the lattice by their cores sizes
$r~\gtrsim~aL$. Such corrections should vary as $1/\rho^2$, or
$1/\sqrt{a^4 V}$
\cite{Smith:1998wt}.

We see from Fig.~\ref{fig_zero_vol} that the data at fixed lattice
spacing follows this approximate volume scaling quite closely.  We
might ask whether $\la \Lambda_{0} \ra$ really goes to zero in the
large volume limit at finite lattice spacing. Fits to our data
allowing a non-zero intercept are not stable, but a large reduction in
statistical errors, as well as simulations in larger volumes, would be
needed to answer this question definitively.

\subsection{Scaling of the low-lying non-zero modes}

We turn now to the low-lying non-zero modes. From the requirement of a
physical spectral density in Eq.~(\ref{eqn_spectral_density}) and the
Banks-Casher relation, Eq.~(\ref{eqn_banks_casher}), it is clear that
the size of these is governed, up to a constant, by the product of the
bare chiral condensate and the volume $\Sigma V$. Specifically, the
number of modes in a small interval $(0, \delta)$ should scale as
$\sim \delta \rho(0) V \sim \delta \Sigma V$, and therefore we expect
each mode to scale as $(\Sigma V) ^{-1}$. The variation of these modes
with changing lattice spacing and volume is therefore fixed by the
response of the bare chiral condensate.

\begin{figure}[t]
\includegraphics[width=3in, clip]{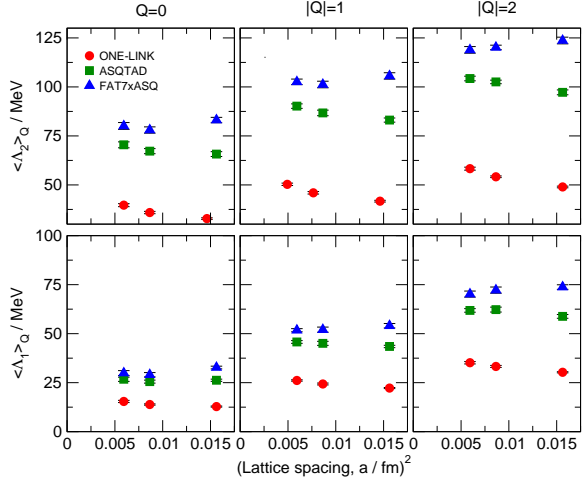}

\caption{\label{fig_mean_scal} 
  The scaling of the non-zero modes with lattice spacing at fixed lattice
  volume for different topological charge sectors. In each case the
  quartets are averaged. }
\end{figure}

In Fig.~\ref{fig_mean_scal} we show the variation of the first two
non-zero mode quartets $\la \Lambda_{1,2} \ra$ with lattice spacing at
fixed physical volume ($aL \approx 1.5$~fm) for several $Q$. For all
the improved actions the changes are very small, indicating only small
discretization errors, as for the chiral condensate. For the near-zero
modes, the deviations from the leading order scaling behaviour seem to
be of opposite sign for \asqtad\ and \hisq.  The \naive\ data show
much larger deviations, of the same sign as for the \asqtad\ data.

The different actions appear to lack a common continuum limit. This is
not surprising: as discussed above, the plot is inversely related to
the density and therefore the bare chiral condensate.  As explained
earlier, this receives a large renormalisation for the \naive\ action
compared to the improved actions that means that the bare results are
not comparable.

\begin{figure}[b]
\includegraphics[width=3in, clip]{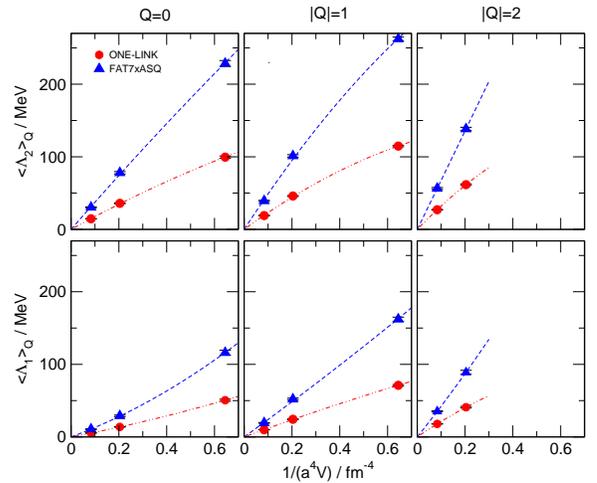}

\caption{\label{fig_mean_vol} 
  The scaling of the non-zero modes with volume at fixed lattice
  spacing for different topological charge sectors. In each case the
  quartets are averaged. Also shown are linear plus quadratic fits
  through the origin.}
\end{figure}

From the arguments above, we expect $\la \Lambda_s \ra$ to vary as the
inverse volume at fixed lattice spacing. We find this to be so for
both \naive\ and improved actions.  Fig.~\ref{fig_mean_vol} shows the
results for $s=1,2$.

Given the different volume scaling of the two sets of eigenmodes, we
cannot really talk about a gap between near-zero and non-zero modes:
we can always choose a volume large enough that the ``non-zero'' modes
(suppressed as $1/V$) are numerically smaller than the ``near-zero''
modes (which fall only as $1/\sqrt{V}$). The modes are still
qualitatively different however, and can be separated by their
chirality.

\begin{figure}[t]
\includegraphics[width=3in, clip]{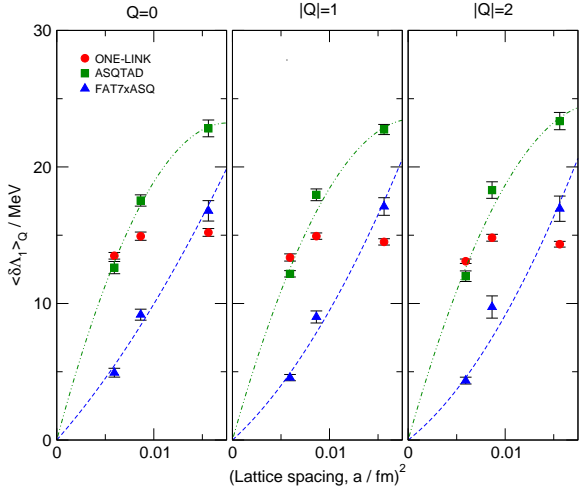}

\caption{\label{fig_split_scal} The scaling of the first non-zero
  intra-quartet splitting with lattice spacing at fixed lattice volume
  in different topological charge sectors.  Higher quartets have
  smaller splittings. The lines are linear plus quadratic fits for
  \asqtad\ and \hisq\ . }
\end{figure}

Taste breaking interactions disappear in the continuum limit and the
eigenmodes should recover a 4-fold degeneracy. We plot the
intra-quartet splitting of the first quartet, $\la \delta \Lambda_1
\ra$ for different sectors of topological charge in
Fig.~\ref{fig_split_scal} (higher quartets have smaller splittings).
The lattice volume is fixed at $aL \approx 1.5$~fm. The splitting
falls with lattice spacing, as expected, and in a way consistent with
the $\mathcal{O}(a^2)$ dependence of the taste breaking
interactions. The splitting depends only weakly on the topological
charge.

This is the only quantity we have studied where there is a difference
of more than one standard error between \hyp\ and \hisq\ results; the
\hisq\ results are marginally closer to our continuum expectations.
(This was also true in Fig.~\ref{fig_zero_scal}, but the difference
was not statistically significant).

There are strong similarities between this plot and
Fig.~\ref{fig_zero_scal}: the improved fermion actions all show only
very weak deviations from the naive scaling behaviour; the
$\mathcal{O}(a^4)$ corrections are positive for \hisq\ and negative
for \asqtad; the \naive\ results however show very large deviations
from the expected behaviour. Numerically, the \hisq\ splittings are
smaller than the \naive\ for $a \lesssim 0.12$~fm; the \asqtad\
numbers are less for $a \lesssim 0.08$~fm.

\begin{figure}[t]
\includegraphics[width=3in, clip]{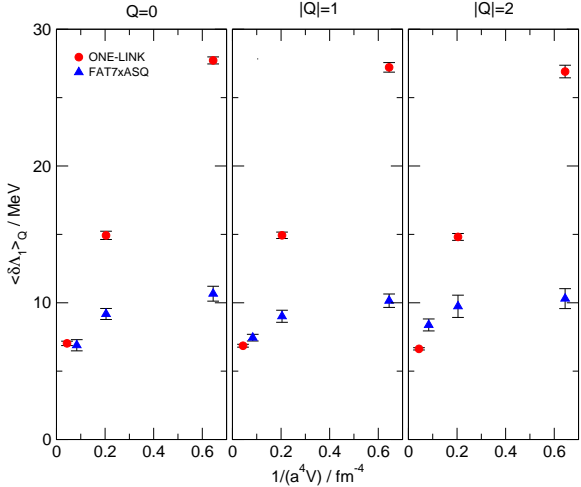}

\caption{\label{fig_split_vol} The scaling of the first non-zero
  intra-quartet splitting with volume at fixed lattice spacing in
  different topological charge sectors. Higher quartets have smaller
  splittings.  See comments in Fig.~\ref{fig_zero_scal}. }
\end{figure}

In Fig.~\ref{fig_split_vol} we show the variation of the splitting
with volume at fixed lattice spacing. Once more the splitting is
independent of the topological charge. There is a striking difference
in the volume dependence: the \naive\ splittings vary markedly whilst
the improved action results are nearly insensitive.

\begin{figure}[t]
\includegraphics[width=3in, clip]{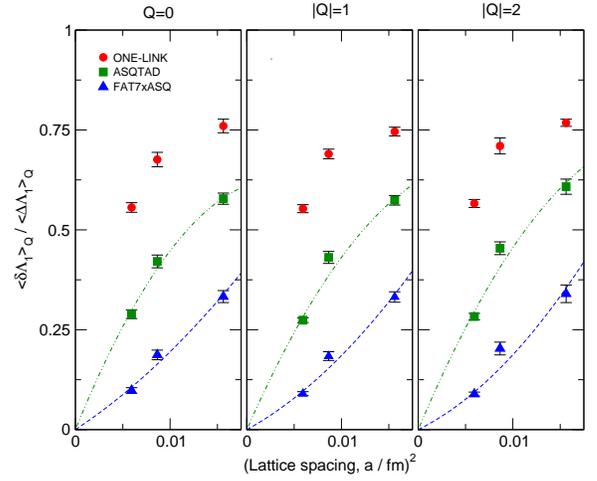}

\caption{\label{fig_gap_rat_scal} The ratio of the intra-quartet
  splitting of the first non-zero quartet to the inter-quartet gap
  between first and second quartets as a function of lattice spacing
  at fixed lattice size $aL \approx 1.5$~fm. The lines are linear plus
  quadratic fits for \asqtad\ and \hisq\ .}
\end{figure}

To satisfy ourselves that there is a clear restoration of the taste
symmetry, we need the intra-quartet splitting $\la \delta \Lambda_s
\ra$ to be much less than the gap between neighbouring quartets $\la
\Delta \Lambda_s \ra$. In Fig.~\ref{fig_gap_rat_scal} we show the
ratio of these as a function of lattice spacing for $s=1$ (where it is
largest).

Again, the ratios show little dependence on the topological charge. In
all cases the improved action results are better than the \naive. This
is increasingly true at small lattice spacings, as the improved
actions show a ratio falling clearly to zero as $a^2$.

We can understand these results quite easily. As the quartets become
degenerate, $\la \Delta \Lambda_s \ra$ will tend towards the
difference between the quartet means. Fig.~\ref{fig_mean_scal} shows
this is controlled by the chiral condensate and has little lattice
spacing dependence for the improved actions. The intra-quartet
splitting falls as $a^2$ and so, therefore, will the ratio.

\section{RMT Predictions}
\label{sec_rmt}

\begin{figure}[t]
\includegraphics[width=3in,clip]{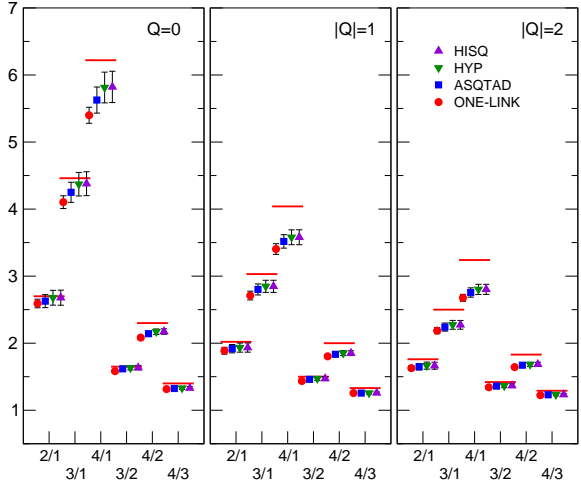}

\caption{\label{fig_ratios_acts} The ratios of expectation values of
  small eigenvalues for the $aL = 1.5$~fm, $a=0.093$~fm ensemble for
  different Dirac operators, compared with the predictions based on a
  universal distribution (horizontal lines) for $|Q| \le 2$.}
\end{figure}
\begin{figure}[t]
\includegraphics[width=3in,clip]{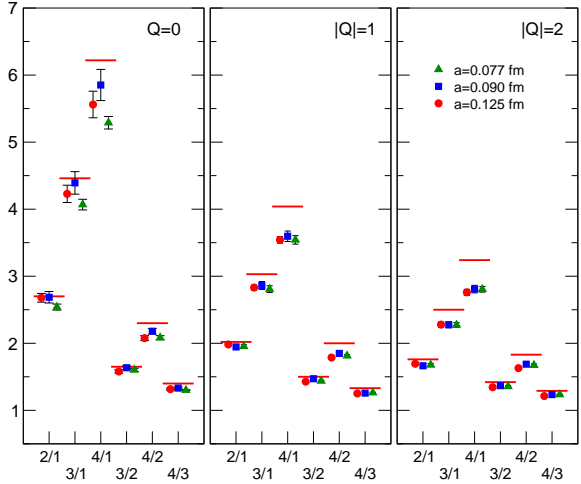}

\caption{\label{fig_ratios_scal} The ratios of expectation values of
  small eigenvalues for the $aL = 1.5$~fm ensembles for the \hisq\
  operator compared with the predictions based on a universal
  distribution (horizontal lines) for $|Q| \le 2$ .}
\end{figure}

As discussed previously, the chiral symmetries of staggered quarks are
more complicated than for continuum QCD. At finite lattice spacing the
$N_t^2=16$ pions split into 5 multiplets containing $(1,4,6,4,1)$
states, and only one of these states becomes massless in the chiral
limit. This is known as the ``Goldstone pion'', with mass $M_G$.  The
fifteen remaining states have masses $M_{\text{NG}}$ arising from the
taste breaking interactions. The (chirally extrapolated) masses are
$\mathcal{O}(a^2)$ and therefore zero only in the continuum limit.

There are potentially \textit{two} universal regions for different
parameters of the system:
\begin{eqnarray}
\mbox{$\varepsilon$-regime:} ~~~ &&
(\Lambda_{\text{QCD}})^{-1} \ll L \ll (M_{\text{NG}})^{-1} \; ,
\nonumber \\
\mbox{$\varepsilon^\prime$-regime:} ~~~ &&
(M_{\text{NG}})^{-1} \ll L \ll (M_{\text{G}})^{-1} \; .
\nonumber
\end{eqnarray}
The first corresponds to Eq.~(\ref{eqn_rmt_hier}), with the same
chiral symmetries that we expect in the continuum.  The low-lying
non-zero modes have a near $N_t$-fold degeneracy, and follow the
predictions of Eqs.~(\ref{eqn_rmt_all},\ref{eqn_rmt_k}). These are
\textit{exactly the same} distributions as for continuum QCD (and
chiral lattice fermions
\cite{Edwards:1999ra,Edwards:1999zm,Damgaard:1999tk,
Hasenfratz:2002rp,Bietenholz:2003mi,Giusti:2003gf}).

At finite lattice spacing there is a second,
$\varepsilon^\prime$-regime, where the finite volume partition
function describes only the static mode of the single Goldstone
pion. There is not even approximate restoration of the continuum
symmetries, and the associated RMT has only $U(1) \otimes U(1)$ chiral
symmetry.  The universal predictions for the eigenvalues are therefore
strikingly different from those for continuum QCD. In particular, the
predictions are the same for all sectors of topological charge
\cite{Damgaard:PrivComm}.
Presumably it is this regime that was studied in
\cite{Berbenni-Bitsch:1998tx,Damgaard:1998ie,
  Gockeler:1998jj,Damgaard:1999bq,Damgaard:2000qt}.
With the coarse lattices and unimproved gauge ensembles used in these
studies, they observed no sensitivity of the eigenvalue spectra to
$Q$, leading to the incorrect folklore that staggered quarks are
``blind to the topology''.

To see the continuum chiral symmetries, there must be a sufficiently
large mass gap between the heaviest pion and the lightest of the other
hadrons. This requires the use of improved gauge and fermion actions
and a sufficiently small lattice spacing. In particular if we increase
the size of the lattice $L$, the lattice spacing must be reduced
accordingly in order to satisfy
$L~\ll~(M_{\text{NG}})^{-1}$. Otherwise we would start to see the
effects of the mass of the non-Goldstone pions.  At coarse lattice
spacings, with unimproved fermions, $M_{\text{NG}} \approx
\Lambda_{\text{QCD}}$ and there will be no $\varepsilon$-regime.

If the lattice spacing is keep fixed and we increase the volume, we
expect there will be a non-universal crossover until the lattice
reaches the $\varepsilon^\prime$-regime when universal behaviour will
resume. Of course, it would be interesting to simulate a range of
lattice sizes and witness the crossover between the two universal
regions.  The range of lattice spacings is, however, well beyond the
resources of this study.  We instead concentrate on the continuum-like
region.

\subsection{Comparison}
\label{sec_epsilon}

We begin by comparing the ratios of the mean non-zero eigenvalues
$\left\langle \Lambda_s \right\rangle_Q / \left\langle \Lambda_t
\right\rangle_Q$ (which we denote ``s/t'') to the ratios from
RMT. Factors of the chiral condensate scale away, so this provides a
parameter--free test of the $\varepsilon$-regime predictions.

\begin{figure}[t]
\includegraphics[width=3in,clip]{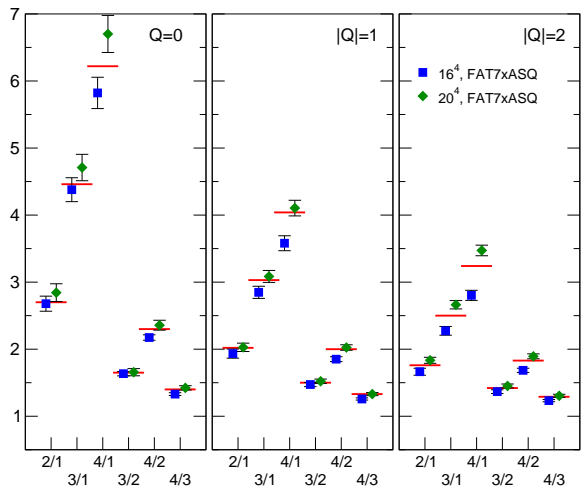}

\caption{\label{fig_ratios_vol} The ratios of expectation values of
  small eigenvalues of the \hisq\ operator for two $a = 0.093$~fm
  ensembles of different volumes compared with the predictions based
  on a universal distribution (horizontal lines) for $|Q| \le 2$.}
\end{figure}

We show the eigenvalue ratios in
Figs.~\ref{fig_ratios_acts}-\ref{fig_ratios_vol}. The ratios differ
markedly between different topological charge sectors, showing clearly
that the staggered fermions are in no way ``topology blind''. Also
shown are the ratios predicted by RMT. Fig.~\ref{fig_ratios_acts}
shows that all the staggered actions agree quite well with the
expected ratios at $a = 0.093$~fm.  We therefore have good evidence
for the existence of a continuum $\varepsilon$-regime. It also implies
that the continuum chiral symmetries are present, with the correct
pattern of spontaneous breaking.

There are some deviations from the universal predictions. They are not
discretization effects: Fig.~\ref{fig_ratios_scal} shows the \hisq\
ratios for a range of different lattice spacings at fixed volume $aL
\approx 1.5$~fm and discrepancies exist even on the finest
lattice. Increasing the system size from $1.5$ to $1.8$~fm in
Fig.~\ref{fig_ratios_vol} does reduce the size of the disagreement.
Indeed, on the larger volume the ratios agree with RMT at least as
well as for chiral fermions on comparable lattices
\cite{Giusti:2003gf}.

Fig.~\ref{fig_ratios_acts} appears to suggest that the \naive\ Dirac
operator fits the universal predictions as well as the improved
operators. This appears so even on the coarsest lattices despite the
large pion splitting and lack of obvious spectral quartets in the
\naive\ spectrum.

In this respect the ``ratios plots'' can be misleading. Firstly they
only compare the mean eigenvalues with RMT and provide no information
as to whether the shape of the individual distributions match. Also,
when comparing with RMT we must rescale all the eigenvalues using the
chiral condensate. Agreement with the ratio plot does not imply that
eigenvalues in different topological charge sectors are scaled by the
same factor.

\begin{figure*}[t]
\includegraphics[width=7in,clip]{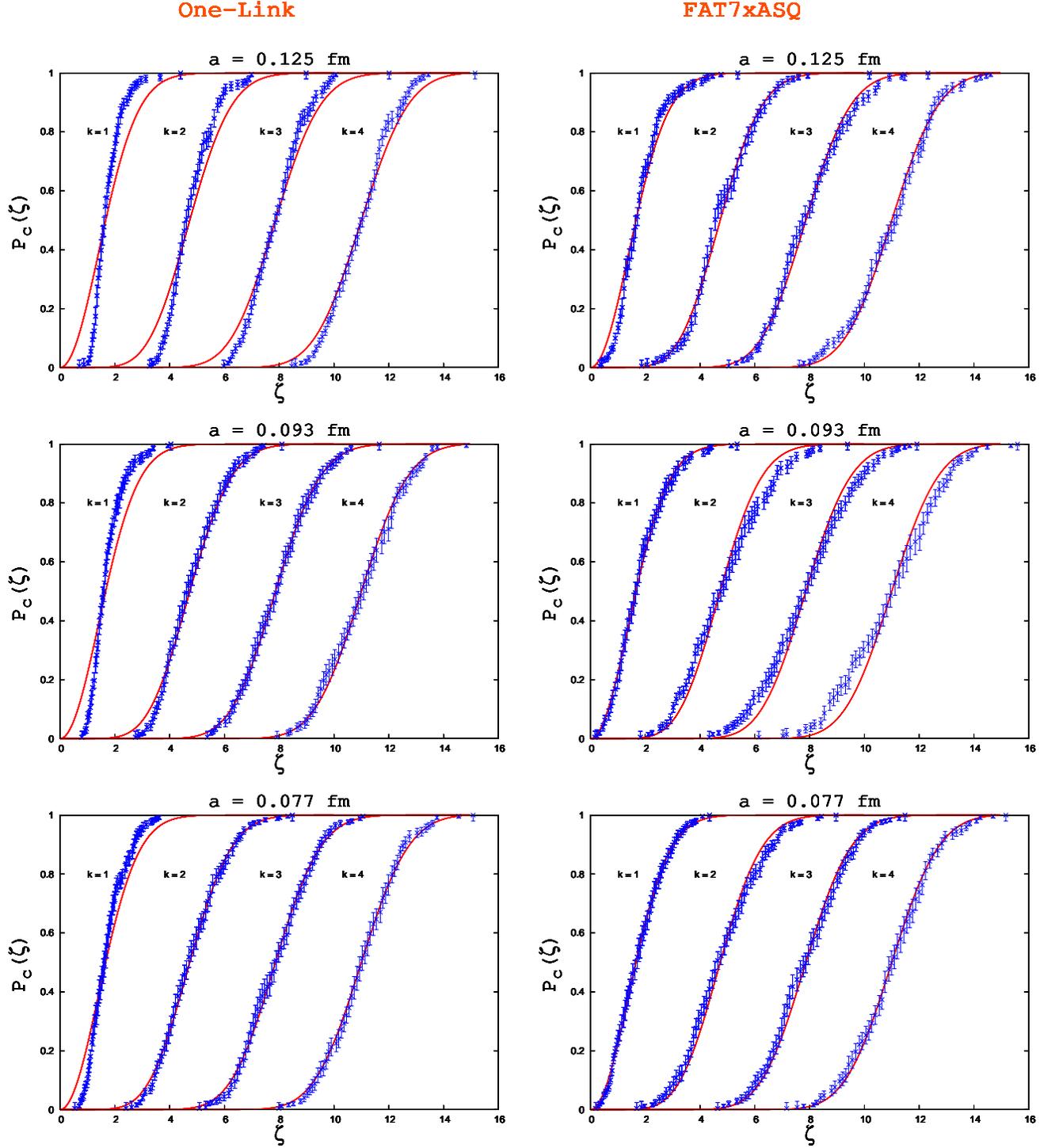}

\caption{\label{fig_rmt_fits_Q0} Comparison of the unfolded eigenvalue
  distribution with RMT for the lowest eigenvalues in the $Q=0$
  sector. The lattice size is kept at $aL \approx 1.5$~fm}
\end{figure*}

\begin{figure*}[t]
\includegraphics[width=7in,clip]{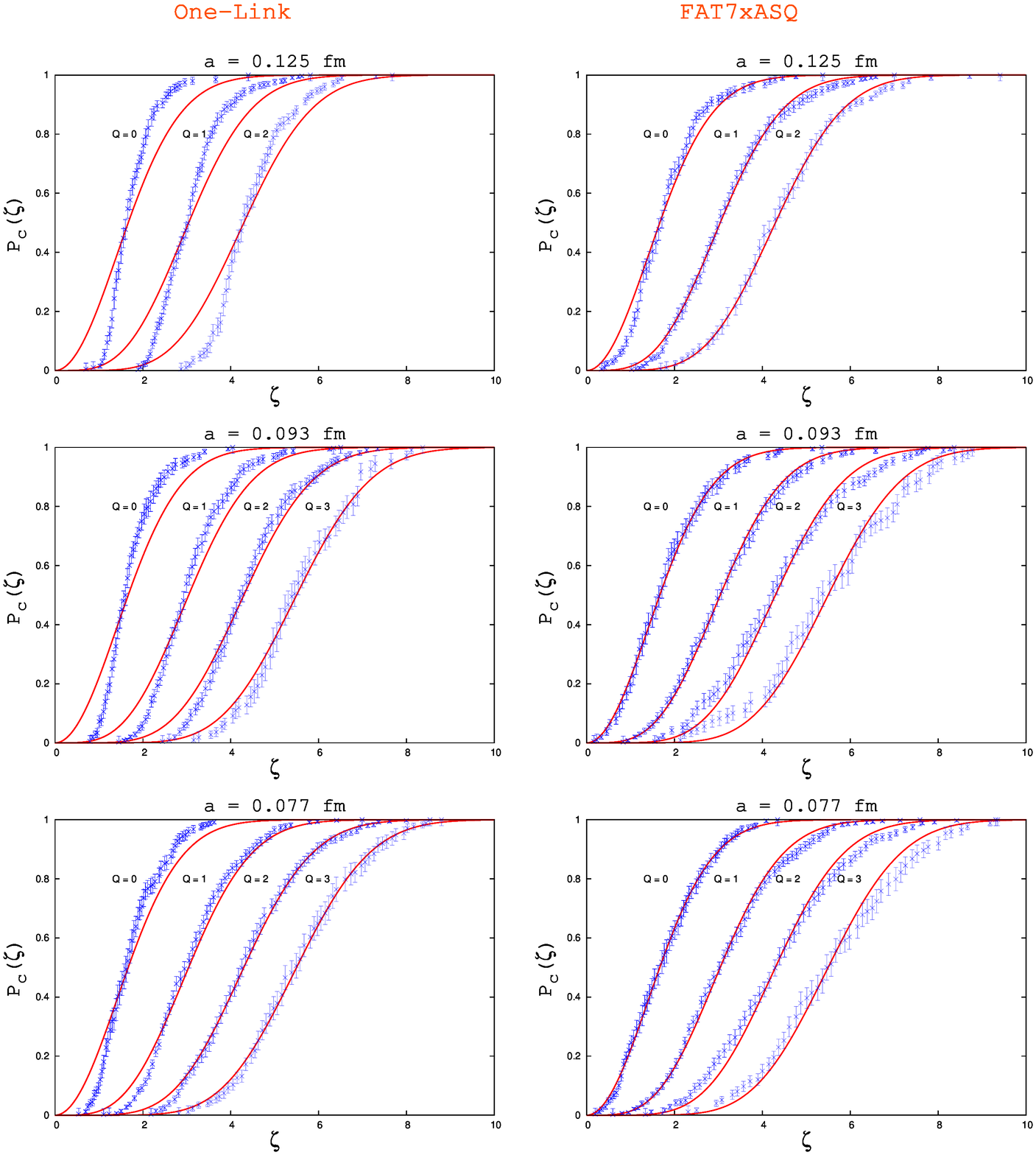}

\caption{\label{fig_rmt_fits_k1} Comparison of the unfolded eigenvalue
  distribution with RMT for the lowest eigenvalue for different
  topological charges. The lattice size is kept at $aL \approx 1.5$~fm}
\end{figure*}

To address the first point we separately fit the individual spectral
densities to the predictions from RMT. These one parameter fits based
on Eq.~(\ref{eqn_unfolding}) yield a prediction for the chiral
condensate. To avoid the ambiguities of histogram bin-size we use the
cumulative eigenvalue distributions,
\begin{equation}
p(\lambda) = \int_0^\lambda d\lambda^\prime \; \rho(\lambda^\prime) \; .
\end{equation}
We show examples of the fits in Figs.~\ref{fig_rmt_fits_Q0}
and~\ref{fig_rmt_fits_k1}.

For all operators the fits are worst for low $s$ and for small $|Q|$.
We can qualitatively understand this in terms of eigenvalue repulsion.
The taste changing interactions have split the near-zero modes, which
has a knock-on effect on the rest of the spectrum as the eigenvalues
avoid level crossings.  The lower quartets are closest to the
near-zero modes and are thus most strongly affected. As the quartets
get numerically larger as we increase the topological charge, so the
repulsion should be less.

It is clear that the improved Dirac operators match the predictions of
RMT very closely over the full range of lattice spacings. By contrast,
even at $a=0.077$~fm the \naive\ Dirac operator still shows
significant discrepancies.

Separately fitting the $(Q=0, k=1..4)$ and $(|Q|=0..3, k=1)$
eigenvalue distributions to the RMT predictions gives 7 values for the
bare chiral condensate for each Dirac operator on each ensemble.  We
find these numbers to be broadly consistent. The spread in the numbers
is greatest for the \naive\ action and on the smallest volumes. This
is not surprising, when such eigenvalue distributions showed the
greatest deviation from the RMT predictions

The lower half of Table~\ref{tab_rmt_chir_cond} gives estimates for
the chiral condensate obtained from the median of the seven individual
numbers.  The quoted errors are half the range, which forms the
dominant uncertainty.

We can compare these with the model-independent estimates of $\Sigma$
obtained using the Banks--Casher relation. There is very good
agreement between the numbers. The dominant source of error is
different in each case: for the Banks-Casher relation it comes from
the choice of fit range and function used to extrapolate the spectral
density to zero eigenvalue. For the RMT comparisons, it comes from the
variation in fit parameters for different eigenvalues, and is
therefore related to exactly how well the effective partition function
describes the lattice theory. The good agreement of the numbers
suggests both systematic biases are under control. See section
\ref{sec_Banks-Casher} for a discussion of how the bare chiral
condensate must be renormalised to extract a physical result.

\section{Conclusions and Outlook}
\label{sec_summary}

It is widely believed that any ``problems'' with a given lattice Dirac
operator will show themselves most clearly in the topologically
sensitive, low-lying eigenmodes. In this study we have studied these
modes for the staggered lattice Dirac operator, using a range of
improved and unimproved versions. The improved operators pass all the
tests. In particular, and contrary to previous accepted wisdom, such
fermions do respond exactly as expected to the gluonic topological
charge, and in a way identical to continuum QCD and other lattice
formulations of the Dirac operator.

We have seen that for improved operators the eigenvalue spectrum
divides cleanly into near-zero and non-zero modes. The near-zero modes
are characterised by a uniformly high chirality, with their relative
number fixed by the Atiyah-Singer Index Theorem. Indeed they could be
used to define the topological charge $Q$ of a configuration. Their
eigenvalues scale with lattice spacing and volume as expected.

The non-zero modes, by contrast, have chirality that is near zero.
They divide into near degenerate quartets, with the splitting reducing
to zero as $a^2$ in common with the taste breaking interactions. The
quartet means are controlled by the chiral condensate and scale with
lattice spacing and volume accordingly. In addition, the low-lying
non-zero modes follow closely the universal distributions predicted by
random matrix theory.

Since the near-zero and non-zero modes scale differently with the
volume, it is clear that the ``gap'' between the highest near-zero and
the lowest non-zero modes is a volume dependent quantity. At finite
lattice spacing we can always go to a volume large enough that the two
sets of eigenvalues have comparable magnitude. This is not really an
issue because the gap is not a physically observable quantity: what
matters is the size of the near-zero modes compared to the light quark
masses and whether there is a well--defined Index Theorem. As we have
seen the two sets of modes are distinguishable by their chirality,
even in the large volume limit.

A quantitative study of axial anomaly physics will require the
near-zero modes to be smaller than the light quark masses,
i.e. $\lambda < \cal{O}({\rm 5 MeV})$, or, at least, smaller than the
sea and valence quark masses used in the simulation. Then the
eigenvalue spectrum is cut off at the low end by the quark mass and
the fact that the near-zero modes are not actually zero is irrelevant.

To understand what this means for our quenched ensembles, it is
necessary to define the size of the typical near-zero mode.  The mean
zero mode $\langle \lambda_0 \rangle_Q$ increases as we increase $Q$,
due to eigenvalue repulsion. For our $\beta=4.8$, $L=20$ lattice we
find that the mean zero-mode is fitted quite well by $\langle
\lambda_0 \rangle_Q \approx (3 + Q)$~MeV for $Q=1..3$.  If we define
the typical topological charge $\overline{Q}$ to be the RMS value from
the topological susceptibility, $\overline{Q} = \sqrt{\chi V} = 2.6$,
then the typical near-zero mode is $\overline{\lambda}_0 = (3 + 2.6)
\approx 6$~MeV. This suggests that we are already able to study the
quenched axial anomaly.

We can attempt to extrapolate our quenched results to ask whether this
condition is true on existing configurations from the MILC
collaboration
\cite{Aubin:2004wf}.
This requires various unsupported, if reasonable, assumptions, not
least that sea quark effects do not strongly affect the size of
individual near-zero modes and that $\langle \lambda_0 \rangle_Q$
continues to be linear in $Q$.

Typical dynamical simulation volumes are larger than we have used in
this quenched study, to avoid finite volume effects in hadron
spectroscopy. As well as an overall $\sqrt{V}$ suppression in the size
of the near-zero modes, this will also affect the typical value
$\overline{Q}$ and therefore the amount of eigenvalue repulsion. There
are two competing effects: the sea quark induced screening of
topological charge reduces $\chi$, whilst the larger volume increases
the width of the histogram of $Q$.

Consider the $a=0.09$~fm MILC ensemble with sea quark masses
$m_s/m_{u,d} = 5$ and volume $V=28^3 \times 96$. Using the data in
\cite{Bernard:2003gq},
we obtain $\overline{Q} = 7.3$. Assuming the quenched relation above
holds, we have \mbox{$\overline{\lambda}_0 = (3 + 7.3) / \sqrt{28^4
/20^4} \sim~6$~MeV} (to be conservative we assume that the volume
scaling of the near-zero modes goes with the smallest length of the
lattice, therefore $28^4$ instead of $28^3 \times 96$). This is very
encouraging. The lightest sea quark mass presently used in these
dynamical simulations is $m_0 = 16$~MeV, which is several times larger
than this projected near-zero mode.

Even given the caveats surrounding this rough estimate, it does seem
likely that a systematic study of the QCD axial anomaly is possible on
these configurations and various aspects of this work are now in
progress.

An on-going debate surrounds the theoretical justification of the
methods use to simulate $N_f < N_t$ flavours of sea quarks
\cite{Bunk:2004br,Bunk:2004kf,Hart:2004sz}.
Recent work suggests there is no ambiguity
\cite{Adams:2004mf,Adams:2004wp,Maresca:2004me}.
Our results (and related work elsewhere
\cite{Durr:2004as,Durr:2004rz})
indicates that with improved staggered quarks the spectrum matches
very closely the continuum expectations with clear restoration of
taste symmetry. It therefore paves the way for the full analysis of
the staggered taste basis that is necessary to finally resolve this
issue.

To clarify the role of topology and the issue of the
$\varepsilon$-regime for staggered quarks is important because
improved staggered fermions are currently the only lattice formulation
that offers the prospect of a comprehensive study of the QCD axial
anomaly free from the systematic effects of lattice spacing and
unphysically large sea quark mass.

\begin{acknowledgments}
  
This research is part of the EU integrated infrastructure initiative
hadron physics project under contract number RII3-CT-2004-506078.
  
We thank: Ph.~de~Forcrand for his topological charge measurement code;
A.~Hasenfratz for help in implementing the \hyp\ operator;
G.P.~Lepage, P.~Damgaard and G.~Akemann for useful discussions. E.F.,
C.T.H.D. and Q.M. are supported by PPARC and A.H. by the U.K.  Royal
Society.  The eigenvalue calculations were carried out on computer
clusters at Scotgrid and the Dallas Southern Methodist University. We
thank David Martin and Kent Hornbostel for assistance.

\end{acknowledgments}

\bibliographystyle{h-physrev4}
\bibliography{paper}

\begin{thebibliography}{10}

\bibitem{Follana:2004sz}
E.~Follana, A.~Hart and C.~T.~H. Davies,
\newblock Phys. Rev. Lett. {\bf 93}, 241601 (2004), [hep-lat/0406010].
%%CITATION = HEP-LAT 0406010;%%

\bibitem{Follana:2004wp}
E.~Follana,
\newblock Nucl. Phys. Proc. Suppl. {\bf 140}, 141 (2005), [hep-lat/0409062].
%%CITATION = HEP-LAT 0409062;%%

\bibitem{Wong1}
K.~Y. Wong and R.~M. Woloshyn,
\newblock hep-lat/0407003.
%%CITATION = HEP-LAT 0407003;%%

\bibitem{Wong2}
K.~Y. Wong and R.~M. Woloshyn,
\newblock Phys. Rev. {\bf D71} (2005), [hep-lat/0412001].
%%CITATION = HEP-LAT 0412001;%%

\bibitem{Atiyah:1963}
M.~Atiyah and I.~Singer,
\newblock Bull. Amer. Math. Soc. {\bf 69}, 422 (1963).

\bibitem{Atiyah:1968}
M.~Atiyah and I.~Singer,
\newblock Ann. Math. {\bf 87}, 596 (1968).

\bibitem{Damgaard:2001ep}
P.~H. Damgaard,
\newblock Nucl. Phys. Proc. Suppl. {\bf 106}, 29 (2002), [hep-lat/0110192].
%%CITATION = HEP-LAT 0110192;%%

\bibitem{Akemann:2003tv}
G.~Akemann and P.~H. Damgaard,
\newblock Phys. Lett. {\bf B583}, 199 (2004), [hep-th/0311171].
%%CITATION = HEP-TH 0311171;%%

\bibitem{Leutwyler:1992yt}
H.~Leutwyler and A.~Smilga,
\newblock Phys. Rev. {\bf D46}, 5607 (1992).
%%CITATION = PHRVA,D46,5607;%%

\bibitem{Shuryak:1993pi}
E.~V. Shuryak and J.~J.~M. Verbaarschot,
\newblock Nucl. Phys. {\bf A560}, 306 (1993), [hep-th/9212088].
%%CITATION = HEP-TH 9212088;%%

\bibitem{Nishigaki:1998is}
S.~M. Nishigaki, P.~H. Damgaard and T.~Wettig,
\newblock Phys. Rev. {\bf D58}, 087704 (1998), [hep-th/9803007].
%%CITATION = HEP-TH 9803007;%%

\bibitem{Damgaard:2000ah}
P.~H. Damgaard and S.~M. Nishigaki,
\newblock Phys. Rev. {\bf D63}, 045012 (2001), [hep-th/0006111],
\newblock see corrections at hep-th/0006111v2.
%%CITATION = HEP-TH 0006111;%%

\bibitem{Verbaarschot:1993pm}
J.~J.~M. Verbaarschot and I.~Zahed,
\newblock Phys. Rev. Lett. {\bf 70}, 3852 (1993), [hep-th/9303012].
%%CITATION = HEP-TH 9303012;%%

\bibitem{Forrester1993}
P.~Forrester,
\newblock Nucl. Phys. B {\bf 402}, 709 (1993).

\bibitem{Golterman:1984}
M.~Golterman and J.~Smit,
\newblock Nucl. Phys. {\bf B245}, 61 (1984).

\bibitem{Smit:1987fn}
J.~Smit and J.~C. Vink,
\newblock Nucl. Phys. {\bf B286}, 485 (1987).
%%CITATION = NUPHA,B286,485;%%

\bibitem{Hands:1990wc}
S.~J. Hands and M.~Teper,
\newblock Nucl. Phys. {\bf B347}, 819 (1990).
%%CITATION = NUPHA,B347,819;%%

\bibitem{Venkataraman:1998yj}
L.~Venkataraman and G.~Kilcup,
\newblock Nucl. Phys. Proc. Suppl. {\bf 63}, 826 (1998), [hep-lat/9710086].
%%CITATION = HEP-LAT 9710086;%%

\bibitem{Hasenfratz:2003}
A.~Hasenfratz,
\newblock ``{H}ow good is the {HYP} staggered action?'',
\newblock Talk presented at Lattice 2003 (Tsukuba, Japan).

\bibitem{Follana:2003}
E.~Follana {\em et~al.},
\newblock Nucl. Phys. Proc. Suppl. {\bf 129}, 384 (2004).

\bibitem{Berbenni-Bitsch:1998tx}
M.~E. Berbenni-Bitsch, S.~Meyer, A.~Schafer, J.~J.~M. Verbaarschot and
  T.~Wettig,
\newblock Phys. Rev. Lett. {\bf 80}, 1146 (1998), [hep-lat/9704018].
%%CITATION = HEP-LAT 9704018;%%

\bibitem{Damgaard:1998ie}
P.~H. Damgaard, U.~M. Heller and A.~Krasnitz,
\newblock Phys. Lett. {\bf B445}, 366 (1999), [hep-lat/9810060].
%%CITATION = HEP-LAT 9810060;%%

\bibitem{Gockeler:1998jj}
M.~Gockeler, H.~Hehl, P.~E.~L. Rakow, A.~Schafer and T.~Wettig,
\newblock Phys. Rev. {\bf D59}, 094503 (1999), [hep-lat/9811018].
%%CITATION = HEP-LAT 9811018;%%

\bibitem{Damgaard:1999bq}
P.~H. Damgaard, U.~M. Heller, R.~Niclasen and K.~Rummukainen,
\newblock Phys. Rev. {\bf D61}, 014501 (2000), [hep-lat/9907019].
%%CITATION = HEP-LAT 9907019;%%

\bibitem{Damgaard:2000qt}
P.~H. Damgaard, U.~M. Heller, R.~Niclasen and K.~Rummukainen,
\newblock Phys. Lett. {\bf B495}, 263 (2000), [hep-lat/0007041].
%%CITATION = HEP-LAT 0007041;%%

\bibitem{Hasenfratz:2001wd}
A.~Hasenfratz,
\newblock Phys. Rev. {\bf D64}, 074503 (2001), [hep-lat/0104015].
%%CITATION = HEP-LAT 0104015;%%

\bibitem{Bernard:2002sa}
MILC, C.~Bernard {\em et~al.},
\newblock Nucl. Phys. Proc. Suppl. {\bf 119}, 991 (2003), [hep-lat/0209050].
%%CITATION = HEP-LAT 0209050;%%

\bibitem{Bernard:2003gq}
MILC, C.~Bernard {\em et~al.},
\newblock Phys. Rev. {\bf D68}, 114501 (2003), [hep-lat/0308019].
%%CITATION = HEP-LAT 0308019;%%

\bibitem{Aubin:2004qz}
MILC, C.~Aubin {\em et~al.},
\newblock hep-lat/0409051.
%%CITATION = HEP-LAT 0409051;%%

\bibitem{Kogut:1990qd}
J.~B. Kogut, D.~K. Sinclair and M.~Teper,
\newblock Nucl. Phys. {\bf B348}, 178 (1991).
%%CITATION = NUPHA,B348,178;%%

\bibitem{Bitar:1991wr}
HEMCGC, K.~M. Bitar {\em et~al.},
\newblock Phys. Rev. {\bf D44}, 2090 (1991).
%%CITATION = PHRVA,D44,2090;%%

\bibitem{Kuramashi:1993mv}
Y.~Kuramashi, M.~Fukugita, H.~Mino, M.~Okawa and A.~Ukawa,
\newblock Phys. Lett. {\bf B313}, 425 (1993).
%%CITATION = PHLTA,B313,425;%%

\bibitem{Alles:2000cg}
B.~Alles, M.~D'Elia and A.~Di~Giacomo,
\newblock Phys. Lett. {\bf B483}, 139 (2000), [hep-lat/0004020].
%%CITATION = HEP-LAT 0004020;%%

\bibitem{Durr:2003xs}
S.~Durr and C.~Hoelbling,
\newblock Phys. Rev. {\bf D69}, 034503 (2004), [hep-lat/0311002].
%%CITATION = HEP-LAT 0311002;%%

\bibitem{Durr:2004as}
S.~Durr, C.~Hoelbling and U.~Wenger,
\newblock Phys. Rev. {\bf D70}, 094502 (2004), [hep-lat/0406027].
%%CITATION = HEP-LAT 0406027;%%

\bibitem{Durr:2004ab}
S.~Durr and C.~Hoelbling,
\newblock hep-lat/0408039.
%%CITATION = HEP-LAT 0408039;%%

\bibitem{Durr:2004rz}
S.~Durr, C.~Hoelbling and U.~Wenger,
\newblock hep-lat/0409108.
%%CITATION = HEP-LAT 0409108;%%

\bibitem{Durr:2004ta}
S.~Durr and C.~Hoelbling,
\newblock Phys. Rev. {\bf D71}, 054501 (2005), [hep-lat/0411022].
%%CITATION = HEP-LAT 0411022;%%

\bibitem{Lepage:1998vj}
G.~P. Lepage,
\newblock Phys. Rev. {\bf D59}, 074502 (1999), [hep-lat/9809157].
%%CITATION = HEP-LAT 9809157;%%

\bibitem{Bernard:1999xx}
MILC, C.~W. Bernard {\em et~al.},
\newblock Phys. Rev. {\bf D61}, 111502 (2000), [hep-lat/9912018].
%%CITATION = HEP-LAT 9912018;%%

\bibitem{Orginos:1999cr}
MILC, K.~Orginos, D.~Toussaint and R.~L. Sugar,
\newblock Phys. Rev. {\bf D60}, 054503 (1999), [hep-lat/9903032].
%%CITATION = HEP-LAT 9903032;%%

\bibitem{Naik:1986bn}
S.~Naik,
\newblock Nucl. Phys. {\bf B316}, 238 (1989).
%%CITATION = NUPHA,B316,238;%%

\bibitem{Knechtli:2000ku}
F.~Knechtli and A.~Hasenfratz,
\newblock Phys. Rev. {\bf D63}, 114502 (2001), [hep-lat/0012022].
%%CITATION = HEP-LAT 0012022;%%

\bibitem{Follana:2004}
E.~Follana {\em et~al.},
\newblock Further improvements to staggered quarks on the lattice,
\newblock in preparation.

\bibitem{Curci}
G.~Curci, P.~Menotti and G.~Paffuti,
\newblock Phys.Lett. {\bf B130}, 205 (1983).

\bibitem{Curcierratum}
{\bf Erratum:}, G.~Curci, P.~Menotti and G.~Paffuti,
\newblock Phys.Lett. {\bf B135}, 516 (1984).

\bibitem{Luscher}
M.~Luscher and P.~Weisz,
\newblock Commun. Math. Phys. {\bf 97}, 59 (1985).
%%CITATION = CMPHA,97,59;%%

\bibitem{Luscher_erratum}
{\bf Erratum:}, M.~Luscher and P.~Weisz,
\newblock Commun. Math. Phys. {\bf 98}, 433 (1985).
%%CITATION = CMPHA,97,59;%%

\bibitem{Alford}
M.~Alford, W.~Dimm, G.~Lepage, G.~Hockney and P.~Mackenzie,
\newblock Phys. Lett. {\bf B361}, 87 (1995), [hep-lat/9507010].
%%CITATION = HEP-LAT 0208018;%%

\bibitem{Bonnet:2001rc}
F.~D.~R. Bonnet, D.~B. Leinweber, A.~G. Williams and J.~M. Zanotti,
\newblock Phys. Rev. {\bf D65}, 114510 (2002), [hep-lat/0106023].
%%CITATION = HEP-LAT 0106023;%%

\bibitem{Zhang:2001fk}
J.~B. Zhang {\em et~al.},
\newblock Phys. Rev. {\bf D65}, 074510 (2002), [hep-lat/0111060].
%%CITATION = HEP-LAT 0111060;%%

\bibitem{Davies:2003ik}
HPQCD, C.~T.~H. Davies {\em et~al.},
\newblock Phys. Rev. Lett. {\bf 92}, 022001 (2004), [hep-lat/0304004].
%%CITATION = HEP-LAT 0304004;%%

\bibitem{deForcrand:1995qq}
P.~de~Forcrand, M.~Garcia~Perez and I.-O. Stamatescu,
\newblock Nucl. Phys. Proc. Suppl. {\bf 47}, 777 (1996), [hep-lat/9509064].
%%CITATION = HEP-LAT 9509064;%%

\bibitem{deForcrand:1997sq}
P.~de~Forcrand, M.~Garcia~Perez and I.-O. Stamatescu,
\newblock Nucl. Phys. {\bf B499}, 409 (1997), [hep-lat/9701012].
%%CITATION = HEP-LAT 9701012;%%

\bibitem{Cullum_book}
J.~Cullum and R.~A. Willoughby,
\newblock {\em Lanczos Algorithms for Large Symmetric Eigenvalue Computations.
  Vol 1, theory} (SIAM Society for Industrial \& Applied Mathematics,
  Philadephia, 2002).

\bibitem{Hein}
J.~Hein, Q.~Mason, G.~Lepage and H.~Trottier,
\newblock Nucl. Phys. Proc. Suppl. {\bf 106} (2002), [hep-lat/0110045].
%%CITATION = HEP-LAT 0110045;%%

\bibitem{Trottier:2002}
H.~Trottier, G.~Lepage, P.~Mackenzie, Q.~Mason and M.~Nobes,
\newblock Nucl. Phys. Proc. Suppl. {\bf 106} (2002), [hep-lat/0110147].
%%CITATION = HEP-LAT 0110147;%%

\bibitem{Lee:2002}
W.~Lee and S.~Sharpe,
\newblock Phys. Rev. {\bf D66}, 114501 (2002), [hep-lat/0208018].
%%CITATION = HEP-LAT 0208018;%%

\bibitem{Aubin:2004mas}
C.~Aubin {\em et~al.},
\newblock Phys. Rev. {\bf D70}, 031504 (2004), [hep-lat/0405022].
%%CITATION = HEP-LAT 0405022;%%

\bibitem{Mason}
HPQCD and UKQCD, Q.~Mason {\em et~al.},
\newblock Phys. Rev. Lett., in press , [hep-lat/0503005].
%%CITATION = HEP-LAT 0503005;%%

\bibitem{Giusti:1998}
L.~Giusti, F.~Rapuano, M.~Talevi and A.~Vladikas,
\newblock Nucl. Phys. {\bf B538}, 249 (1999), [hep-lat/9807014].
%%CITATION = HEP-LAT 9807014;%%

\bibitem{Smith:1998wt}
UKQCD, D.~A. Smith and M.~J. Teper,
\newblock Phys. Rev. {\bf D58}, 014505 (1998), [hep-lat/9801008].
%%CITATION = HEP-LAT 9801008;%%

\bibitem{Edwards:1999ra}
R.~G. Edwards, U.~M. Heller, J.~E. Kiskis and R.~Narayanan,
\newblock Phys. Rev. Lett. {\bf 82}, 4188 (1999), [hep-th/9902117].
%%CITATION = HEP-TH 9902117;%%

\bibitem{Edwards:1999zm}
R.~G. Edwards, U.~M. Heller, J.~E. Kiskis and R.~Narayanan,
\newblock Phys. Rev. {\bf D61}, 074504 (2000), [hep-lat/9910041].
%%CITATION = HEP-LAT 9910041;%%

\bibitem{Damgaard:1999tk}
P.~H. Damgaard, R.~G. Edwards, U.~M. Heller and R.~Narayanan,
\newblock Phys. Rev. {\bf D61}, 094503 (2000), [hep-lat/9907016].
%%CITATION = HEP-LAT 9907016;%%

\bibitem{Hasenfratz:2002rp}
P.~Hasenfratz, S.~Hauswirth, T.~Jorg, F.~Niedermayer and K.~Holland,
\newblock Nucl. Phys. {\bf B643}, 280 (2002), [hep-lat/0205010].
%%CITATION = HEP-LAT 0205010;%%

\bibitem{Bietenholz:2003mi}
W.~Bietenholz, K.~Jansen and S.~Shcheredin,
\newblock JHEP {\bf 07}, 033 (2003), [hep-lat/0306022].
%%CITATION = HEP-LAT 0306022;%%

\bibitem{Giusti:2003gf}
L.~Giusti, M.~Luscher, P.~Weisz and H.~Wittig,
\newblock JHEP {\bf 11}, 023 (2003), [hep-lat/0309189].
%%CITATION = HEP-LAT 0309189;%%

\bibitem{Damgaard:PrivComm}
P.~H. Damgaard,
\newblock private communication, 2004.

\bibitem{Aubin:2004wf}
C.~Aubin {\em et~al.},
\newblock hep-lat/0402030.
%%CITATION = HEP-LAT 0402030;%%

\bibitem{Bunk:2004br}
B.~Bunk, M.~Della~Morte, K.~Jansen and F.~Knechtli,
\newblock Nucl. Phys. {\bf B697}, 343 (2004), [hep-lat/0403022].
%%CITATION = HEP-LAT 0403022;%%

\bibitem{Bunk:2004kf}
B.~Bunk, M.~Della~Morte, K.~Jansen and F.~Knechtli,
\newblock hep-lat/0408048.
%%CITATION = HEP-LAT 0408048;%%

\bibitem{Hart:2004sz}
A.~Hart and E.~M\"uller,
\newblock Phys. Rev. {\bf D70}, 057502 (2004), [hep-lat/0406030].
%%CITATION = HEP-LAT 0406030;%%

\bibitem{Adams:2004mf}
D.~H. Adams,
\newblock hep-lat/0411030.
%%CITATION = HEP-LAT 0411030;%%

\bibitem{Adams:2004wp}
D.~H. Adams,
\newblock Nucl. Phys. Proc. Suppl. {\bf 140}, 148 (2005), [hep-lat/0409013].
%%CITATION = HEP-LAT 0409013;%%

\bibitem{Maresca:2004me}
F.~Maresca and M.~Peardon,
\newblock hep-lat/0411029.
%%CITATION = HEP-LAT 0411029;%%

\end{thebibliography}

\end{document}